\documentclass[a4paper,11pt]{article}
\pdfoutput=1 

\usepackage{jheppub} 
\usepackage[dvipsnames]{xcolor}
\usepackage[T1]{fontenc} 

\usepackage{subcaption}

\title{\boldmath Complexity and Operator Growth for Quantum Systems in Dynamic Equilibrium}

\author[a]{Cameron Beetar,}
\author[a]{Nitin Gupta}
\author[a,b]{S. Shajidul Haque,}
\author[a,b]{Jeff Murugan,}
\author[a,b]{and Hendrik J R Van Zyl,}

\affiliation[a]{The Laboratory for Quantum Gravity \& Strings, Department of Mathematics and Applied Mathematics, University of Cape Town, Cape Town, South Africa}
\affiliation[b]{The National Institute for Theoretical and Computational Sciences, Private Bag X1, Matieland, South Africa}

\abstract{Krylov complexity is a measure of operator growth in quantum systems, based on the number of orthogonal basis vectors needed to approximate the time evolution of an operator. In this paper, we study the Krylov complexity of a $\mathsf{PT}$-symmetric system of oscillators, which exhibits two phase transitions that separate a dissipative state, a Rabi-oscillation state, and an ultra-strongly coupled regime. We use a generalization of the $su(1,1)$ algebra associated to the Bateman oscillator to describe the Hamiltonian of the coupled system, and construct a set of coherent states associated with this algebra. We compute the Krylov (spread) complexity using these coherent states, and find that it can distinguish between the $\mathsf{PT}$-symmetric and $\mathsf{PT}$ symmetry-broken phases. We also show that the Krylov complexity reveals the ill-defined nature of the vacuum of the Bateman oscillator, which is a special case of our system. Our results demonstrate the utility of Krylov complexity as a tool to probe the properties and transitions of $\mathsf{PT}$-symmetric systems. }

\begin{document} 
\maketitle
\flushbottom

\section{Introduction and Motivation}
\label{sec:intro}
One of the grand challenges in quantum physics is a complete understanding of how quantum systems reach dynamic equilibrium; a state where the macroscopic observables are stationary, but the microscopic dynamics are still non-trivial. Such dynamic equilibria are ubiquitous in nature, and  are realized in a range of settings, such as thermal baths \cite{Sachdev:2011fcc,Eisert:2014jea,Heyl:2018jzi}, black holes \cite{Wald:1997qz,Wald:1999vt,Barbon:2003aq}, or even the everyday physics of saturated solutions \cite{PhysRevLett.63.1070}. A key feature of dynamic equilibrium is that it involves a detailed balance between entropy production and dissipation which, in turn, implies that the system is not isolated, but coupled to an environment, in a very specific way. Environmental issues or, more precisely, the physics of open quantum systems are neither new nor unexplored, seeing as how they are crucial to decoherence, dissipation, and noise in any realistic quantum system. However, the field has seen a resurgence of activity of late, in the context of the time-development, or growth, of operators \cite{Haque:2021kdm,Bhattacharyya:2021fii,Bhattacharyya:2020iic,Schuster:2022bot,Bhattacharyya:2022rhm,Bhattacharjee:2022lzy,Bhattacharya:2022gbz,Liu:2022god,NSSrivatsa:2023qlh,Bhattacharya:2023zqt} and the relation thereof to quantum chaos and information scrambling \cite{Ali:2019zcj,Bhattacharyya:2020kgu,Bhattacharyya:2020rpy,Bhattacharyya:2020art,Bhattacharyya:2020qtd,Bhattacharjee:2022vlt}.\\

\noindent
Operator growth \cite{Parker:2018yvk,Magan:2020iac,Yin:2020oze,Patramanis:2021lkx,Hornedal:2022pkc} is a phenomenon that characterizes the computational complexity and chaos properties of quantum systems. It refers to the fact that a simple operator, such as a local observable or a unitary gate, can evolve into a highly non-local and entangled operator under the action of a Hamiltonian. The degree of operator growth can be quantified by various measures, such as the out-of-time-ordered correlator (OTOC) \cite{Larkin1969QuasiclassicalMI}, the frame potential \cite{Roberts:2016hpo}, or the Krylov complexity \cite{Rabinovici:2020ryf, Barbon:2019wsy, Dymarsky:2019elm,Caputa:2021sib,Balasubramanian:2022tpr}. The latter is a measure of operator growth based on the Krylov subspace, itself the linear span of the successive powers of a matrix applied to a vector. Krylov complexity of an operator is then defined as the number of orthogonal basis vectors needed to approximate its time evolution to within a given accuracy and can be seen as a generalization of the circuit complexity \cite{Chapman:2021jbh}, which counts the number of elementary gates needed to implement a unitary operator. The Krylov complexity has the advantage of being applicable to any operator, not just unitary ones, and of being independent of the choice of the gate set. Moreover, the Krylov complexity can be computed quite efficiently using numerical methods such as the Lanczos algorithm \cite{Lanczos1950AnIM} and has been used to study the operator growth in various quantum systems, such as the Sachdev-Ye-Kitaev (SYK) model \cite{Roberts:2018mnp,Jian:2020qpp,Bhattacharjee:2022ave,Bhattacharjee:2023uwx}, the Heisenberg spin chain \cite{PhysRevE.104.034112,Rabinovici:2021qqt,Rabinovici:2022beu}, and conformal field theories \cite{Caputa:2021ori,Dymarsky:2021bjq,Kundu:2023hbk}. \\

\noindent
The Lanczos coefficients that are central to the Lanczos algorithm and, by extension, the Krylov complexity are actually encoded in the thermal Wightman 2-point function \cite{Parker:2018yvk}. Specifically, given an operator $O$,  the associated Krylov complexity $K_{O}(t)$ is determined by the spectral density of the commutator of the operator at different times via
\begin{eqnarray}
    K_O(t) = \frac{1}{2} \int_{-\infty}^{\infty} d\omega \frac{1}{e^{\beta \omega} - 1} \left| \langle [O(t), O(0)] \rangle_\omega \right|^2\,,
\end{eqnarray}
where $O(t)$ is the corresponding Heisenberg operator, $\beta$ is the inverse temperature, and $\langle \cdot \rangle_\omega$ is the Fourier transform of the thermal expectation value. This relation between the Krylov complexity and the thermal two-point function can, in turn, be used to probe novel properties of quantum systems such as phase transitions and critical points. Equipped with this tool then, we cycle back to our original motivation and ask: {\it what is the simplest system that exhibits a quantum dynamical equilibrium?} And, as usual when trying to answer such questions, the harmonic oscillator is a pretty good place to start. In this case, we take as a prototype two linearly coupled oscillators, one damped and one anti-damped; a system that manifests in, for example, whispering-gallery-mode optical resonators \cite{Bender:2013qta,Peng:2014idi}. With the coupling turned off, this system reduces to the famed {\it Bateman model} of a damped harmonic oscillator \cite{PhysRev.38.815}.\\

\noindent
As described in detail in \cite{Bender:2013qta}, this system of two coupled oscillators possesses a remarkably rich phase space. Depending on the strength of the coupling and the relative strength of the damping coefficients, it exhibits two phase transitions that separate a dissipative state (by which we will mean uni-directional energy flow from one oscillator to the other), a Rabi-oscillation state, and an ultra-strongly coupled regime. In this sense, the oscillatory regime of the coupled oscillators provides a tractable example of a dynamic equilibrium without the added complications that obscure more realistic quantum systems such as spin chains or some SYK variants. The price that we pay, of course, is that the Hilbert space of   bosonic oscillators is infinite dimensional. As we will see shortly, this adds non-trivial, but not insurmountable, complications to the numerical methods we use. Even though the system is in general non-Hermitian\footnote{Though the Hamiltonian is naively hermitian w.r.t. the standard Hilbert space inner product it is unbounded outside of the Rabi oscillation phase.}, in its loss/gain-balanced Rabi phase, it is $\mathsf{PT}$-symmetric. Hamiltonians in this class are characterised by the fact that they are non-Hermitian, but invariant under the combined action of parity ($\mathsf{P}$) and time-reversal ($\mathsf{T}$) symmetries\footnote{The reader may well have encountered such systems before under the name $\mathcal{PT}$-symmetric systems; we just don't like calligraphic script.}. They have attracted substantial attention in recent years, due to their intriguing properties and applications in various fields of physics, such as optics, acoustics, quantum mechanics, and quantum field theory \cite{El-Ganainy:2018ksn,Ozdemir2019, Ashida:2020dkc}. For our purposes, the most important of these properties is that fact that the combined $\mathsf{PT}$ symmetry renders the spectrum of the non-Hermitian Hamiltonian real. In this article, we compute and study the Krylov complexity of this system of oscillators to understand (\textit{i}) how to distinguish between the $\mathsf{PT}$-symmetric and $\mathsf{PT}$ symmetry-broken phases of matter and (\textit{ii}) how operators grow in and around the $\mathsf{PT}$-symmetric dynamic equilibrium phase.\\

\noindent
The idea behind Krylov complexity is rooted in the mathematics of matrix diagonalization algorithms. Specifically, given a Hamiltonian, $H$, the Krylov complexity for a particular initial operator $O_{0}$ is computed by
\begin{itemize}
    \item constructing the Krylov subspace, which is the linear span of the operators $O_0, H O_0$, $H^2 O_0, \dots, H^n O_0$, obtained by applying the Hamiltonian repeatedly to the initial operator,
    \item orthonormalizing the Krylov subspace using, for example, the Gram-Schmidt procedure. The orthonormalized Krylov subspace is a set of operators $O_0, O_1, \dots, O_n$, such that $\langle O_i, O_j \rangle = \delta_{ij}$, where $\langle \cdot, \cdot \rangle$ is some inner product on the space of operators, such as the Hilbert-Schmidt inner product,
    \item approximating the time evolution of the initial operator by a linear combination of the orthonormalized Krylov operators. The approximation is given by $O(t) \approx \sum_{i=0}^n c_i(t) O_i$, where $c_i(t)$ are some coefficients that depend on the time and the Hamiltonian and, 
    \item calculating the Krylov complexity $K_O(t) = \min_{ n \in \mathbb{N}} \left| O(t) - \sum_{i=0}^n c_i(t) O_i \right| < \epsilon$ of the time-evolved operator by counting the number of orthonormalized Krylov operators that are needed to achieve a desired accuracy. 
\end{itemize}

\noindent 
This algorithm can be further simplified if, in addition, the Hamiltonian can be expressed as an element of some (low-rank) symmetry algebra such as the Heisenberg \cite{Caputa:2021sib} or Jacobi algebras \cite{Haque:2022ncl}. Key to this simplification are the coherent states associated to the algebra \cite{Perelomov,Gazeau:2009zz}, that form an overcomplete basis for the Hilbert space of the system, and in terms of which any state or operator can be expanded. Instead of applying the Hamiltonian repeatedly to the initial operator, which can be a high-dimensional matrix, one can apply the a ladder operator repeatedly to an initial coherent state, which is a low-dimensional vector. The Krylov subspace is then the linear span of the coherent states obtained by applying the annihilation operator to the initial coherent state. Apart from this computational advantage, the use of coherent states also reveals some physical insights into the operator growth and Krylov complexity, by mapping them to classical motions in phase space \cite{Provost:1980nc, Ashtekar:1997ud, Brody:1999cw}. Here, coherent states can be represented as points, and the annihilation operator can be represented as a vector field that generates a flow on the symplectic manifold. The operator growth can then be seen as the displacement of the coherent state in phase space, and the Krylov complexity can be seen as the volume of the region spanned by the Krylov state \cite{Caputa:2021sib, Chattopadhyay:2023fob}.\\

\noindent
The algebraic representation of the (quantum) Bateman oscillator dates back to \cite{osti_6577304} where the formulation of the Bateman Hamiltonian  in terms of $su(1,1)$ elements played a crucial role in its solution. In this article, we use a generalization of this $su(1, 1)$ symmetry to give an algebraic formulation of the $\mathsf{PT}$-symmetric Hamiltonian of the coupled system, and use this to construct a set of coherent states associated with this algebra. We then compute the Krylov (or, more precisely, the Spread) complexity using the algorithm described above. To summarise our results; we find that the Krylov complexity can indeed distinguish between the $\mathsf{PT}$-symmetric and $\mathsf{PT}$ symmetry-broken phases. Moreover, the behaviour of the Krylov complexity in the appropriate limit reveals the ill-defined nature of the vacuum of the Bateman oscillator, as a special case of our system. Finally, we will conclude by drawing some universal lessons from this particular problem about the nature of operator growth in more general systems with balanced loss and gain.\\

\section{A Review of the Bateman Oscillator}

We begin by unpacking the details of the system we will be studying in this paper, largely following a discussion in \cite{Bender:2013qta}.

\subsection{Bateman Oscillator}

\label{sec:Bateman}
\noindent 
Consider the simple damped harmonic oscillator whose displacement from equilibrium $x(t)$ satisfies,
\begin{equation}
    \label{dampedSHO}
    \ddot{x}(t) + 2\gamma \dot{x}(t) + \omega^2 x(t) = 0\,, 
\end{equation}
with the mass of the oscillator $m=1$. This system is clearly not unitary since energy is dissipated. The question then, is how to  {\it quantize} it, since conventional quantization approaches (canonical, path integral etc.) generally require a Lagrangian for which \eqref{dampedSHO} is the equation of motion. To address this question, Bateman \cite{PhysRev.38.815} introduced an auxiliary oscillator, whose position is, say $y(t)$, is governed by an {\it anti-damped} oscillator equation,
\begin{equation}
    \label{anti-dampedSHO}
    \ddot{y}(t) - 2\gamma \dot{y}(t) + \omega^2 y(t) = 0\,.  
\end{equation}
The Lagrangian of the composite system is then given by,
\begin{equation}
    \label{Bateman}
    L = \dot{x}\dot{y} + \gamma(x\dot{y}-\Dot{x}y) - \omega^2 xy\,.
\end{equation}
Notice that the Euler-Lagrange equation for $x$ is \eqref{anti-dampedSHO} and that for $y$ is the damped oscillator equation \eqref{dampedSHO} (similar to the action for JT gravity). An additional, important point is that \eqref{anti-dampedSHO} is the time-reversed version of \eqref{dampedSHO}. This means that bilinear forms such as the Hamiltonian can be time-independent.
The Hamiltonian can be constructed in the usual way. The momenta associated to the oscillator variables $x(t)$ and $y(t)$ are,
\begin{eqnarray*}
    p_{x} &\equiv& \frac{\partial L}{\partial \Dot{x}} = \Dot{y} - \gamma y   \\
    p_{y} &\equiv& \frac{\partial L}{\partial \Dot{y}} = \Dot{x} + \gamma  x
\end{eqnarray*}
so that, 
\begin{equation}
\label{BatemanNoInt}
    H = p_{x}\Dot{x} + p_{y}\Dot{y} - L = p_{x}p_{y} + \gamma(yp_{y}-xp_{x}) + \varkappa xy \equiv H_{0} + H_{1}\,,
\end{equation}
where we have defined $\varkappa \equiv \omega^2 - \gamma^{2}$, $H_{0} \equiv p_{x}p_{y} + \varkappa xy$ and $H_{1} \equiv \gamma(yp_{y}-xp_{x})$. The position and momentum variables satisfy a canonical Heisenberg algebra,
\begin{eqnarray*}
    [x,p_{x}] = [y,p_{y}] = i\hbar\,,
\end{eqnarray*}
with all other commutators vanishing. Substituting the expressions for the canonically conjugate momenta, these can be written as
\begin{equation*}
    [x,\Dot{y}] = [y,\Dot{x}] = i\hbar\,.
\end{equation*}
While this is not the focus of our problem, one way to solve this system is to replace the position-momentum basis with a creation-annihilation basis. This is done (assuming $\varkappa>0$) by defining the creation and annihilation operators,
\begin{eqnarray*}
    a_x &=& \frac{1}{\sqrt{2\hbar \sqrt{\varkappa} }}\left(p_{x} - i\sqrt{\varkappa} x\right)\,,\quad 
    a_y = \frac{1}{\sqrt{2\hbar \sqrt{\varkappa}}}\left(p_{y} - i\sqrt{\varkappa} y\right)\,,\\
    a_x^{\dagger} &=& \frac{1}{\sqrt{2\hbar \sqrt{\varkappa}}}\left(p_{x} + i\sqrt{\varkappa} x\right)\,,\quad 
    a_y^{\dagger} = \frac{1}{\sqrt{2\hbar \sqrt{\varkappa}}}\left(p_{y} + i\sqrt{\varkappa} y\right)\,,
\end{eqnarray*}
which satisfies the usual algebra,
\begin{eqnarray*}
    \left[a_x,a_{x}^{\dagger}\right] &=& \left[a_y,a_y^{\dagger}\right] = 1\,,\\
    \left[a_x,a_y\right] &=& \left[a_x^{\dagger},a_y^{\dagger}\right] = 0\,.
\end{eqnarray*}
In terms of the creation and annihilation operators the part of the Hamiltonian that survives the $\gamma\to 0$ limit,
\begin{eqnarray*}
    H_{0} = \hbar\sqrt{\varkappa} \left(a_x^{\dagger}a_y + a_y^{\dagger}a_x\right)\,.
\end{eqnarray*}
We can write this in a more suggestive way by writing
\begin{equation}
    A \equiv \frac{a_x+a_y}{\sqrt{2}}\,,\quad
    B \equiv \frac{a_x-a_y}{\sqrt{2}}\,,     
    \label{ABdefinition}
\end{equation}
which satisfy the same algebra as the $a$ and $b$ ladder operators and in terms of which,
\begin{equation}
    \label{H0}
    H_{0} = \hbar\sqrt{\varkappa} \left(A^{\dagger}A - B^{\dagger}B\right)\,.
\end{equation}
while the perturbation Hamiltonian,
\begin{equation}
    \label{H1}
    H_{1} = i\hbar \gamma\left(A^{\dagger}B^{\dagger} - AB\right)\,.
\end{equation}
Some comments are in order:
\begin{enumerate}
    \item Eigenvalues of the number operators $A^{\dagger}A$ and $B^{\dagger}B$ are non-negative integers. This means that the spectrum of $H_{0}$ is of the form $\hbar\sqrt{\varkappa} (n_{A}-n_{B})$ with corresponding eigenstates $|n_{A},n_{B}\rangle$. Since only the difference of integers enters the spectrum, any other state of the form $|n_{A}+n, n_{B}+n\rangle$ with $n\in\mathbb{Z}$, is automatically also an eigenstate with the same eigenvalue.
    \item In the $\gamma\to 0$ limit, the states that match those of the undamped harmonic oscillator are those that satisfy $B|\psi\rangle = 0$. 
    \item Since the perturbative operator $H_{1}$ commutes with $H_{0}$, if we start with an undamped state {\it i.e.} an eigenstate of $H_{0}$ of the form $|n_{A},0\rangle$, the effect of $H_{1}$ is to mix in states of the form $| n_{A}+n_{B}, n_{B}\rangle$. In other words, the sinks into which the energy from the damped oscillator goes are states generated by $B^{\dagger}$.
\end{enumerate}
Now, let's use the $A$ and $B$ operators to construct the following operators,
\begin{eqnarray}
    X&\equiv& \frac{1}{2}\left(A^{\dagger}B^{\dagger} + AB\right)\,,\nonumber\\
    Y&\equiv& \frac{i}{2}\left(A^{\dagger}B^{\dagger} - AB\right)\, \label{epsilon0Alg},\\
    Z&\equiv&\frac{1}{2}\left(A^{\dagger}A + BB^{\dagger}\right)\,.\nonumber
\end{eqnarray}
These operators satisfy the algebra,
\begin{eqnarray}
    \left[X,Y\right] &=& iZ\,,\nonumber\\
    \left[Z,Y\right] &=& iX\,,\\
    \left[X,Z\right] &=& iY\,,\nonumber
\end{eqnarray}
which we identify as the non-compact algebra $su(1,1)$.  We can also write this algebra in the basis of ladder operators 
\begin{equation}
L_{+} = X + Y\,, \ \ \  \ \ \ L_{-} = X - Y\,, \ \ \ \ \ \ L_{0} = Z\,,
\end{equation}
which satisfy
\begin{equation}
\left[ L_{-}, L_{+}  \right] = 2 L_0\,, \ \ \  \ \ \ \left[ L_{0}, L_{\pm}  \right] = 0\,.
\end{equation}
In terms of this algebra, note that
\begin{enumerate}
    \item the perturbation Hamiltonian $H_{1}$ is proportional to the $Y$ operator. Specifically, $H_{1} = i\hbar\gamma Y$, and
    \item $H_{0}$ is related to the Casimir of $su(1,1)$ since,
    \begin{equation}
        Z^{2} - X^{2} - Y^{2} = h_{0}^{2} - \frac{1}{4}\,,
    \end{equation}
    with $2\hbar\sqrt{\varkappa} h_{0} = H_{0}$. By definition, $H_{0}$ commutes with all the elements of the algebra.
\end{enumerate}
To summarize the story so far; the full Hamiltonian for the Bateman oscillator is the sum of the Casimir and one element of $su(1,1)$.

\subsection{Interactions and $\mathsf{PT}$ symmetry}

There are two ways to interpret the Bateman  system. In the first, the $x$-oscillator constitutes the `system' and the $y$-oscillator is regarded as an auxiliary degree of freedom. It exists only to facilitate the variational problem for the dissipative term. In the second, the system is enlarged so that both oscillators are treated as system degrees of freedom. In the latter interpretation, the system with two degrees of freedom is conservative with exactly balanced loss and gain. Consequently, flows in the phase-space of this system are volume-preserving, a feature that persists even in the presence of interactions,
\begin{eqnarray}
    \label{gen-bateman}
    \ddot{x} + 2\gamma\dot{x} +  \omega^2 x + F_{1}(x,y) = 0\,,\nonumber\\
    \\
    \ddot{y} - 2\gamma\dot{y} + \omega^2  y + F_{2}(x,y) = 0\,,\nonumber
\end{eqnarray}
where $F_{1}$ and $F_{2}$ are arbitrary functions, with $F_{1} = F_{2} = 0$ corresponding to the  Bateman oscillator. One remarkable feature of the coupled system \eqref{gen-bateman} is that it is invariant under a combination of spatial parity $\mathsf{P}$ and time-reveral $\mathsf{T}$ symmetries where \cite{PhysRevA.88.062111},
\begin{eqnarray}
    \mathsf{P}&:&x\to -y,\quad y\to -x,\quad p_{x}\to-p_{y},\quad p_{y}\to-p_{x}\,,\\
    \mathsf{T}&:&x\to x,\quad y\to y,\quad p_{x}\to-p_{x},\quad p_{y}\to-p_{y}\,.
\end{eqnarray}
In other words, the parity operator exchanges the loss and gain oscillators while the time-reversal operator sends $t\to-t$. Note, however, that the system is {\it not} symmetric under either $\mathsf{P}$ or $\mathsf{T}$ individually. As such, the generalised Bateman oscillator is an example of a $\mathsf{PT}$-symmetric system \cite{Bender:2013qta}. A hallmark of $\mathsf{PT}$-symmetric systems with balanced loss and gain is that they exhibit phase transitions. To be concrete, let's fix $F_{1}(x,y) = \epsilon y$ and $F_{2}(x,y) = \epsilon x$ with $\epsilon\in\mathbb{R}$ to linearly couple the two oscillators. In this case, there is a simple physical intuition for this. Energy from the $x$-oscillator is lost to, and gained from the $y$-oscillator. If the coupling constant $\epsilon$ is sufficiently small, then energy from the $y$-oscillator does not transfer to the $x$-oscillator sufficiently fast for the system to equilibrate. Once the coupling exceeds some threshold value, say $\epsilon_{\mathrm{crit}}$, energy from the $y$-oscillator flows into the $x$-oscillator sufficiently fast that the system comes to equilibrium. To quantify this at the classical level, let's look for oscillatory solutions of the form $x = x_{0}e^{i\lambda t}$ and $y = y_{0}e^{i\lambda t}$. Substituting into \eqref{gen-bateman} requires the frequency to satisfy the polynomial equation,
\begin{eqnarray}
    \label{char}
    f(\lambda)\equiv \lambda^{4} + \left(4\gamma^{2} - 2 \omega^2 \right)\lambda^{2} + \omega^{4} - \epsilon^{2} = 0\,.
\end{eqnarray}
This quartic polynomial has 4 roots and, depending on $\epsilon$ they are either all real, 2 real and 2 complex, or all complex. The system reaches equilibrium if all the frequencies are real. Any complex frequencies signal exponential growth or decay. Fortunately, since \eqref{char} is quadratic in $\lambda^{2}$, its roots can be computed exactly as,
\begin{eqnarray}
    \label{roots}
    \lambda^{2} = \left(\varkappa-   \gamma^{2} \right)\pm\sqrt{\epsilon^{2}-4\gamma^{2}\varkappa   }\,.
\end{eqnarray}
From here we can see that the roots are all real when the coupling parameter is in the range,
\begin{eqnarray}
    \label{para-range}
    \epsilon_{1}\equiv 2\gamma\sqrt{\varkappa} \leq\epsilon \leq \varkappa + \gamma^2 \equiv\epsilon_{2}
\end{eqnarray}
In other words, when the coupling is smaller than $\epsilon_{1}$, the system too weakly coupled for the $y$-oscillator to transfer energy back to the $x$-oscillator. Similarly, when the system is ultra-strongly-coupled ($\epsilon>\epsilon_{2}$) it also cannot equilibrate, although this regime is less well understood (and also difficult to realise experimentally). In the range \eqref{para-range}, the $\mathsf{PT}$-symmetry is unbroken and the system is in a dynamic equilibrium that manifests as {\it Rabi oscillations} between the $x$ and $y$ oscillators.\\

\noindent
With the loss and gain parameters equal, the generalised Bateman oscillator is derived from the classical Hamiltonian,
\begin{eqnarray}
    H & = & p_{x}p_{y} + \gamma(y p_{y} -p_{x}x) + \varkappa xy + \frac{\epsilon}{2}\left(x^{2}+y^{2}\right) \nonumber \\
    &=& (p_x + \gamma y)(p_y - \gamma x) + (\varkappa + \gamma^2) x y + \frac{\epsilon}{2}(x^2 + y^2) - i \gamma  
    \, , \label{gen-BatemanH}
\end{eqnarray}
differing from \eqref{BatemanNoInt} by the interaction term proportional to $\epsilon$.  
\\ \\
This Hamiltonian is clearly a sum of terms quadratic in position and momentum operators.  The algebra can be organised according the commutators 
\begin{eqnarray}
\left[ \frac{1}{2}\left( a^\dag a + a a^\dag \right) , a^\dag         \right] & = & a^\dag     \nonumber \\
\left[ \frac{1}{2}\left( b^\dag b + b b^\dag \right) , b^\dag         \right] & = & b^\dag     \nonumber 
\end{eqnarray}
where 
\begin{eqnarray}
a & = & \frac{x + i p_x}{\sqrt{2}}    \nonumber \\
b & = & \frac{y + i p_y}{\sqrt{2}}    \nonumber
\end{eqnarray}
It is thus a rank-2 algebra consisting of ten generators: the twelve possible quadratic combinations, less $[x,p_x]$ and $[y, p_y]$.  Indeed, it is a real subalgebra of $so(5, \mathbb{C})$.  The coherent states associated with this algebra \cite{Perelomov} form a natural basis in which to discuss complexity.  These are obtained by acting with group elements on a fixed state in the Hilbert space.  As the fixed state it is convenient to select the lowest weight state, $|0, 0 \rangle$, defined as
the eigenstate of 
\begin{equation}
\frac{1}{2}\left( b^\dag b + b b^\dag \right)|0,0\rangle = \frac{1}{2}\left( a^\dag a + a a^\dag \right)|0, 0\rangle = \frac{1}{2} |0, 0\rangle\,,
\end{equation}
and annihilated by the operators 
\begin{equation}
\frac{1}{2} a a |0,0 \rangle = \frac{1}{2} b b |0,0 \rangle = a b |0,0 \rangle = b^\dag a |0,0\rangle = a^\dag b |0,0\rangle = 0\,.   \nonumber
\end{equation}
With this choice of fixed state the generalized coherent states are constructed as
\begin{equation}
|z_a, z_b, z_{ab}) = e^{\frac{z_a}{2} a^\dag a^\dag} e^{\frac{z_b}{2} b^\dag b^\dag    } e^{z_{ab} a^\dag b^\dag} |0, 0\rangle \,,   \label{so5CohState}
\end{equation}
so that the overlap of these coherent states is given by
\begin{eqnarray}
& & (\bar{z}_{a}, \bar{z}_{b}, \bar{z}_{ab}| z_{a}, z_{b}, z_{ab} ) \nonumber \\
&=& \left(   1 - ( |z_{a}|^2 + |z_{b}|^2 + 2 |z_{ab}|^2) + (z_{ab}^2 - z_{a} z_{b})(\bar{z}_{ab}^2 - \bar{z}_{a} \bar{z}_{b}) 
 \right)^{-\frac{1}{2}}\,.    \label{CohS}
\end{eqnarray}
Note that $|0, 0\rangle$ is \textit{not} necessarily the vacuum state of the theory since the $a, b$ operators are not necessarily the ladder operators for the Bateman Hamiltonian. In particular, a scaling transformation
\begin{equation}
e^{-i \frac{\alpha}{2} (x p_x + p_x x)} a \, e^{i \frac{\alpha}{2} (x p_x + p_x x)} = \sqrt{ \frac{e^\alpha}{2}   } x + \frac{i}{\sqrt{2 e^\alpha }}p_x \,    \label{scaleTrans}
\end{equation}
might be necessary to relate it to the correct vacuum. As a specific example, the vacuum state for the $A, B$ oscillators (\ref{ABdefinition}) (in the underdamped regime)  is given by
\begin{equation}
 |0, 0\rangle_{A,B} \equiv e^{-i \frac{\log(\hbar \sqrt{\varkappa} )}{4} (x p_x + p_x x + y p_y + p_y y) }|0, 0\rangle \,.  \nonumber
\end{equation}
Correspondingly, the Hamiltonian (\ref{gen-BatemanH}) scales as
\begin{eqnarray}
& & e^{-\frac{i}{4} \alpha (xp_x + p_x x + y p_y + p_y y )} H e^{\frac{i}{4} \alpha (xp_x + p_x x + y p_y + p_y y )}    \nonumber \\
&=& e^{-\alpha} \left( p_{x}p_{y} + e^{\alpha} \gamma(yp_{y}-xp_{x}) + e^{2\alpha} \varkappa  xy + e^{2\alpha} \frac{\epsilon}{2}\left(x^{2}+y^{2}\right)    \right)\,,    \nonumber 
\end{eqnarray}  
so that the scaling transformation can always be absorbed into a redefinition of the coefficients and timescale.  \\ \\
Before proceeding we note that there has been some debate \cite{Deguchi:2018otx, Deguchi:2019laq, Bagarello:2019yag, Bagarello2020, Fernandez2020} as to whether the Hamiltonian (\ref{gen-BatemanH}) with $\epsilon \rightarrow 0$ possesses a sensible ground state. 
Consider the transformation \cite{Deguchi:2018otx}
 \begin{eqnarray}
 e^{-\theta(A^\dag B^\dag + A B)} A  e^{\theta(A^\dag B^\dag + A B)} & = & \cos\theta A + \sin\theta B^\dag    \nonumber \\
  e^{-\theta(A^\dag B^\dag + A B)} B  e^{\theta(A^\dag B^\dag + A B)} & = & \cos\theta B + \sin\theta A^\dag    \nonumber 
 \end{eqnarray}
 which puts the relevant Hamiltonian into the form 
 \begin{eqnarray}
 & & e^{-\theta(A^\dag B^\dag + A B)} (H_0 + H_1)  e^{\theta(A^\dag B^\dag + A B)}   \nonumber \\
 & = & \sqrt{\varkappa} (A^\dag A - B^\dag B ) -i \gamma \sin(2\theta)(A^\dag A +  B^\dag B + 1) +    i \gamma \cos(2\theta) (A B - A^\dag B^\dag)   \label{HABDiag}
 \end{eqnarray}
This transformation is not unitary and the resulting form is not hermitian with respect to the standard inner product\footnote{The exception is when $\theta$ is a purely imaginary number in which case this is a Bogoliubov transformation.}.    Performing this same similarity transformation on the state $|0, 0\rangle_{AB}$ yields
\begin{equation}
e^{-\theta (A^\dag B^\dag + A B)}|0, 0\rangle_{AB} = \frac{1}{\cos\theta} e^{-\tan\theta  A^\dag B^\dag  } |0, 0\rangle_{AB}
\end{equation}
which we can normalise, provided $\theta \neq \frac{\pi}{4}$ to give
\begin{equation}
|\theta, 0,0\rangle_{AB} \equiv \frac{\cos(\theta + \theta^*)}{ \cos\theta \cos\theta^* } e^{-\tan\theta  A^\dag B^\dag  } |0, 0\rangle_{AB}    \label{thetaState}
\end{equation}
The authors of \cite{Deguchi:2018otx, Deguchi:2019laq} note that the Hamiltonian (\ref{HABDiag}) is in diagonal form when $\theta = \frac{\pi}{4}$, so its spectrum may be computed.  However, the the authors of \cite{Bagarello:2019yag, Bagarello2020, Fernandez2020} have argued that the state (\ref{thetaState}) does not provide a sensible vacuum at $\theta = \frac{\pi}{4}$ since it is not normalizable, so its eigenvalues do not correspond to physically sensible eigenstates. \\ \\
We will comment further on this in due course; for now it is only important to note that  the lowest weight state $|0,0\rangle$ is well-defined in terms of the representation theory of the underlying algebra.  As such, the operator (\ref{gen-BatemanH}) generates a one-dimensional flow contained entirely on this manifold of coherent states.  Whether a sensible vacuum state exists or is contained on the manifold of coherent states is not important for our purposes, since we will only be using these coherent states in our analysis.  Whether the vacuum is contained on the manifold, or not is a point we will return to in section \ref{AnalysisSection}.

\section{Spread Complexity}

Having set up the system, we would now like to 
 study the Hamiltonian (\ref{gen-BatemanH}) in the context of spread complexity \cite{Balasubramanian:2022tpr} and see if and how it encodes the $\mathsf{PT}$-symmetry breaking in the weak-, critical-, and strong-damping regimes respectively.  It is already known that spread complexity can probe topological phase transitions \cite{Caputa:2022eye, Caputa:2022yju} which is characterised by gap closing between the ground state energy and first excitation energy.  It is also known that, in the thermodynamic limit, spread complexity is closely related to the spectral from factor \cite{Erdmenger:2023wjg}.  Since $\mathsf{PT}$-symmetric transitions are also closely related to the spectrum of the Hamiltonian operator, it is reasonable to expect that spread complexity may be able to distinguish these different phases.  We provide here a brief overview of spread complexity and some of its relevant properties.
\\ \\
Spread complexity provides a systematic way to quantify the difficulty of synthesising a desired target state from a given reference state. Given the target state $|\phi_t\rangle$, it is defined as 
\begin{equation}
C( |\phi_t\rangle, \left\{   |B_n\rangle \right\}   ) = \sum_n c_n \langle \phi_t | B_n \rangle \langle B_n | \phi_t \rangle\,,  \label{cosFunc}  
\end{equation}
where $c_n$ is a strictly increasing sequence of numbers and the $|B_n\rangle$ is an ordered basis for the Hilbert space of target states. The $c_n$ is the complexity of each of the basis vectors.  One way to interpret this is that one can synthesize each of the basis vectors with a particular cost, but taking a superposition does not contribute any additional cost.  \\ \\
Given a Hamiltonian, $H$, and a reference state $|\phi_r\rangle$, a natural basis for the Hilbert space is obtained by
\begin{itemize}
    \item acting with powers of the Hamiltonian on this reference state,
    \begin{equation}
       |O_n) = H^n |\phi_r\rangle\,,
    \end{equation}
    \item followed by a Gram-Schmidt orthogonalization,
    \begin{eqnarray}
       |K_0\rangle & = & |\phi_r\rangle\,,   \nonumber \\
       |K_{n+1}) & = & |O_{n+1}) - \sum_{j=0}^n \langle K_n | O_{n+1}) | K_n\rangle\,,  \\
       |K_n\rangle & = & (K_n | K_n)^{-\frac{1}{2}} |K_n)\,.     \nonumber
\end{eqnarray}
\end{itemize}
The resulting set of states $\mathcal{K} = \{|K_{n}\rangle\,:\,n=0,1,2,\ldots\}$ constitute the Krylov basis for the Hilbert space. The naturalness of this basis is due in large part to  the fact that it minimises the cost function (\ref{cosFunc}) of the time-evolved reference state \cite{Balasubramanian:2022tpr}. Practically speaking, the Krylov basis is most easily generated recursively by the Lanczos algorithm \cite{Lanczos1950AnIM, Balasubramanian:2022tpr}
\begin{eqnarray}
|A_{n+1}\rangle & = & (H- a_n) |K_n\rangle - b_n |K_{n-1}\rangle\,,    \nonumber \\
\\
|K_{n}\rangle &=& b_{n}^{-1}|A_{n}\rangle\,,\nonumber
\end{eqnarray}
where the Lanczos coefficients, $a_n  =  \langle K_n | H |K_n\rangle$ and $b_n^2  =  \langle A_n | A_n \rangle$. If, in addition, the return amplitude
\begin{equation}
 R(t) = \langle \phi_r| e^{- i t H} | \phi_r\rangle\,,     \label{ReturnAmplitude}
 \end{equation}
is known, the coefficients can also be computed by the so-called moments method  (see \cite{Dehesa:1981} for more details). In either case, the original Hamiltonian is tri-diagonal when expressed in the Krylov basis, in the sense that
\begin{equation}
H |K_n\rangle = a_n |K_n\rangle + b_{n} |K_{n-1}\rangle + b_{n+1} |K_{n+1} \rangle\,.    \nonumber
\end{equation}   
A pivotal property of spread complexity is that it is invariant under unitary transformations 
\begin{equation}
  C(U |\phi_t\rangle ; U |\phi_r\rangle, U H U^\dag) = C( |\phi_t\rangle ;  |\phi_r\rangle, H )\,,
\end{equation}
a consequence of the fact that the full Krylov basis is related by a unitary transformation {\it i.e.} if $|K_n\rangle$ is the Krylov basis generated by the Hamiltonian, $H$ acting on the reference state $|\phi_r\rangle$, and $U$ is a unitary operator then, 
\begin{equation}
|K_n'\rangle = U |K_n\rangle
\end{equation}
is the Krylov basis generated by the Hamiltonian $U H U^\dag$ acting on the reference state $U |\phi_r\rangle$.  The Lanczos coefficients are invariant under this unitary transformation. An important subset of target states are those obtained by time-evolving the reference state
\begin{equation}
  |\phi_r(t)\rangle = e^{- i t H} |\phi_r\rangle\,,   \label{tTarget}
\end{equation}
and we note that, for this set of states, the equivalence class of Hamiltonians, \begin{equation}
   H' = V H V^\dag \ \ \ \ \textnormal{where} \ \ \ V|\phi_r\rangle = e^{i \phi} |\phi_r\rangle\,   \label{VTrans}
\end{equation}
and $\phi$ is an arbitrary rotation angle, yields identical expressions for the spread complexity. Moreover, for the target state (\ref{tTarget}), the Lanczos algorithm implies the Schrodinger-like equation,
\begin{equation}
  i \partial_t \langle K_n | \phi_r(t)\rangle = a_n \langle K_n | \phi_r(t)\rangle + b_n \langle K_{n-1} | \phi_r(t)\rangle + b_{n+1}\langle K_{n+1} | \phi_r(t)\rangle    \nonumber
\end{equation}
so that, given the set of Lanczos coefficients and return amplitude, the probability amplitudes can be generated recursively as
\begin{eqnarray}
  \langle K_{n+1} | \phi_r(t)\rangle & = & \sum_{m=0}^{n+1} k_{m,n+1} \partial_t^m \langle K_{0} | \phi_r(t)\rangle  \nonumber \\
  k_{m,n+1} & = & \frac{i k_{m-1,n} - a_n k_{m,n} - b_{n} k_{m,n-1}}{b_{n+1}} \label{LanczosProb}
\end{eqnarray}
Consequently, all the information to compute the complexity of a time-evolved reference state is encoded {\it entirely} in the return amplitude (\ref{ReturnAmplitude}).  As such, the computation of this quantity is central to our analysis.  The growth of spread complexity for the time-evolved reference states is determined by the differential equation,
$$ 
  i\frac{\partial}{\partial t} C( 
  |\phi_r(t)\rangle; |\phi_r, H) = \sum_{n} (c_{n+1} - c_n) b_{n+1} \left( \langle K_{n+1} | \phi_r(t)\rangle \langle \phi_r(t)| K_n\rangle  - \langle K_{n} | \phi_r(t)\rangle  \langle \phi_r(t)| K_{n+1}\rangle
  \right)\,,  
$$
and can be interpreted as the rate at which the reference state spreads along the Krylov chain. 
For our purposes, it will suffice to take the weight coefficients $c_n =n$ in the cost function, so that
\begin{equation}
  C(|\phi_t\rangle ; |\phi_r\rangle, H) = \sum_n n \langle \phi_t | K_n\rangle \langle K_n | \phi_t\rangle.  
\end{equation}
and each application of $H$ comes with a unit cost.  The spread complexity with this weight is the average position of the target state along the chain of Krylov basis vectors.  As a final comment, note that the complexity of any eigenstate of the Hamiltonian vanishes trivially. 
 
\subsection{Analytic results}

Closed-form analytic expressions for the probability amplitudes and subsequent spread complexities are possible \cite{Caputa:2021sib, Balasubramanian:2022tpr, Chattopadhyay:2023fob}, especially when the underlying algebra is low dimensional. One such result of particular relevance for us is associated with a return amplitude of the form 
\begin{equation}
R(t) = e^{i(2 c_0 + c h) t} \left( \cosh(\omega t) + i \frac{c}{\omega} \sinh(\omega t)    \right)^{-h}\,,    \label{AnalyticRtForm}
\end{equation}
where $h$ is the value of the highest weight of the representation\footnote{ In our case this is equal to $\frac{1}{4}$ for both species of oscillator.}, {\it i.e.} $K_0 |\phi_r\rangle = h |\phi_r\rangle$.
The Lanczos coefficients associated with this return amplitude are then given by
\begin{eqnarray}
  (b_n)^2 & = & k_1 n  + k_2 n^2\,,     \nonumber \\
  a_n & = & c_0 + c_1 n\,,  \nonumber
\end{eqnarray}
with 
\begin{eqnarray}
  k_1 & = & (\omega^2 + c^2) (h-1)\,,   \omega^2   \nonumber \\
  k_2 & = & (\omega^2 + c^2)\,,    \nonumber \\
  c_1 & = & 2 c \,.   \nonumber
\end{eqnarray}
With this as input, the probability amplitudes are then given by 
\begin{equation}
\langle K_n | \phi_r(t)\rangle = R(t) \sqrt{\frac{(h)_{(n)}}{n!}} 
 \left(\frac{\sqrt{c^2 + \omega^2} )}{ i \omega \coth(\omega t) - c
 }   \right)^n
\end{equation}
where $(x)_{(n)} = \frac{\Gamma(x +n)}{\Gamma(x)}$ is the $n^{\mathrm{th}}$ Pochhammer symbol.  The complexity can be computed as 
\begin{eqnarray}
C(|\phi_r(t)\rangle; |\phi_r\rangle ) &=& h(\omega^2 + c^2) \frac{\sinh^2(\omega t) }{\omega^2}
\end{eqnarray}
This is in complete agreement with the quadratic-plus-linear example of \cite{Muck:2022xfc} with the slight generalisation that $a_n$ can be non-zero.  Thus, cases where $(b_n)^2$ is given by a quadratic polynomial and $a_n$ a linear polynomial can be solved analytically.  Cases with real $\omega$ give rise to exponentially growing, unbounded spread complexity and cases with imaginary $\omega$ to periodic and bounded spread complexity.   These regimes can be identified directly from the Lanczos coefficients \cite{Balasubramanian:2022tpr} by considering the sign of
\begin{equation}
\omega^2 = k_2 - \frac{c_1^2}{4}.  
\end{equation}
which encodes the information about the  large $n$ behavior of the coefficients.

\section{Results}

\label{AnalysisSection}

Our primary goal in this work will be to compute the (spread) complexity of the coupled oscillator system as a function of the damping and coupling parameters. As might be expected, this is a nontrivial exercise which will require significant numerical analysis. However, before tackling the general case, we note that there are a few interesting limits that permit analytic tractability and that will prove quite instructive.

\subsection{Limit of $\epsilon \rightarrow 0$ }

\begin{figure}
\centering
\includegraphics[width=0.6\textwidth]{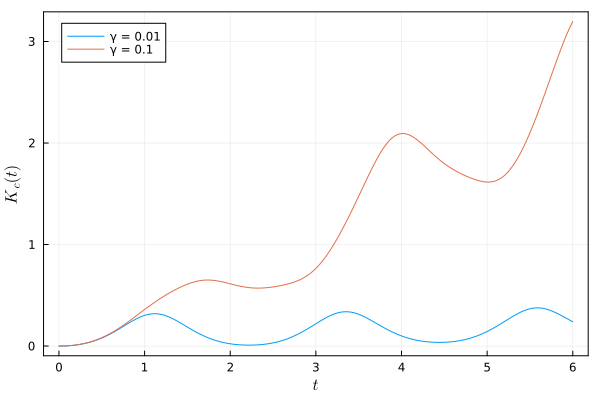}
\caption{The spread complexity of the time-evolved reference state as a function of time with $\varkappa = 2.0,\ \epsilon = 0.0$, and varying values for $\gamma$ for the under-damped weakly coupled system. The reference state is chosen as the coherent state with $\Omega = 1.0$ and $z_{ab} =0$ (throughout the text). The first 50 probability amplitudes are included in the summation, capturing at least 95\% of the probability. This plot illustrates the tension between oscillation and exponential-growth time-scales. Choosing ${ \gamma = 0.01 }$ results in a smaller oscillation time-scale than growth time-scale and hence, the growth is barely perceptible; while for ${ \gamma = 0.1 }$ the opposite is true and the growth in complexity is clear even at early times.}
\label{zeroCouplingPlots}
\end{figure}

As discussed, the generators making up $H'$ are contained in a $u(1,1)$, spanned by the operators in (\ref{epsilon0Alg}).  We can characterise the manifold of coherent states by focussing on the highest weight states.  Analytic results are possible when restricting to coherent states of the form
\begin{equation}
|z\rangle =  (1 - z \bar{z})^{\frac{1}{4}} \ e^{z A^\dag B^\dag} |0,0\rangle_{AB}\,.    \nonumber
\end{equation}
Explicitly, when starting with a coherent state reference state 
\begin{equation}
|z_{ab}\rangle = e^{z_{ab} A^\dag B^\dag} |0,0\rangle_{AB}\,,    \nonumber
\end{equation}
and Hamiltonian $H_1 + H_0$, the spread complexity is given by \cite{Chattopadhyay:2023fob} 
\begin{equation}
C(|z\rangle, |z_{ab}\rangle; H_1+H_0) = \frac{(z - z_{ab})(\bar{z} - \bar{z}_{ab})}{(1 - z_{ab} \bar{z}_{ab})(1 - z \bar{z})   } \,.  \label{su11CompEx1}
\end{equation}
Note that this expression is independent of the choice of Hamiltonian. This is a special feature of cases with an underpinning unit rank algebra. The Hamiltonian does, however, enter when we consider the time-evolved reference state as the target state. These states parametrize some orbit on the manifold; see appendix B. Explicitly, we find that
\begin{equation}
|z_{ab}(t)\rangle = N e^{- i t H_1} e^{z_{ab} A^\dag B^\dag} |0,0\rangle_{AB} =  N e^{\frac{z_0 \cosh(\gamma t) - \sinh(\gamma t)}{\cosh(\gamma t) - z_0 \sinh(\gamma t)} A^\dag B^\dag} |0,0\rangle _{AB}\,,   \nonumber 
\end{equation}
which in turn gives
\begin{equation}
 C( |z_{ab}(t)\rangle; |z_{ab}(0)\rangle, H_1  )  = (1 + \sin^2\phi \sinh^2(2 \rho)) \sinh^2(\gamma t)\,.
\end{equation}
In this last expression, we have parameterised $z_{ab} = e^{i\phi} \tanh \rho$. Note that the rate of exponential growth is determined entirely by the damping constant $\gamma$, while the choice of coherent reference state  affects only the overall constant. Note, also, that the spread complexity is non-trivial for all choices of $z_{ab}$.  This is important, because it implies that the states (\ref{thetaState}) are not eigenstates of the Hamiltonian, in line with the arguments of \cite{Bagarello:2019yag, Bagarello2020, Fernandez2020}.  As such, the ground state is not contained on the manifold of normalisable coherent states.\\ \\
The choice of reference state can be made from the larger set of coherent states (\ref{so5CohState}).  As explained in the previous section, complexity is invariant under simultaneous unitary transformations of the reference state, target state and Hamiltonian.   As such, many unitary transformations yield similar results and we will restrict to a smaller subset of reference states that capture the main features of the physics.  These are states of the form
\begin{equation}
    |z_{ab}, \Omega\rangle = N e^{ -\frac{\log(\Omega)}{4} (a^\dag a^\dag - a a + b^\dag b^\dag - b b )   }e^{z_{ab} a^\dag b^\dag } |0,0\rangle \,.
\end{equation}
Physically, the coefficient $z_{ab}$ controls the number of excitations while $\Omega$ tunes the scale of the oscillators as in (\ref{scaleTrans}). The return amplitude is given by 
\begin{eqnarray}
    R(t) &=& \langle \bar{z}_{ab}, \Omega| e^{\bar{z}_{ab} a b } e^{ \frac{\log(\Omega)}{4} (a^\dag a^\dag - a a + b^\dag b^\dag - b b )   } e^{- i t H_1} |z_{ab},\Omega\rangle   \nonumber \\
    &=& \left(\frac{(\Omega^2 - k)^2}{4 k \Omega^2} (1 + c^2)\sin^2(\sqrt{k} t) + \left( \cosh(\gamma t) +  i c \sinh(\gamma t)  \right)^2   \right)^{-\frac{1}{2}} \,,    \nonumber
\end{eqnarray}
where $$c = -i \frac{z_{ab} - \bar{z}_{ab}}{1 - z_{ab} \bar{z}_{ab} } \,.  $$
This reduces to the case with complexity (\ref{su11CompEx1}) when $\Omega^2 = \varkappa$. Note that, in the limit $\gamma \rightarrow 0$, the return amplitude becomes periodic with a period of $\sqrt{\varkappa}$ if $\varkappa > 0$. When $\Omega^2 = \varkappa$, the system is prepared in an eigenstate of $H_0$. However, if $\Omega^2 \neq \varkappa$, $H_0$ induces an oscillating state. Therefore, in the return amplitude mentioned above, two competing mechanisms are at play. The oscillatory behavior of $H_0$ leads to time-evolved reference states oscillating along the chain of Krylov basis vectors with a bounded average position. On the other hand, $H_1$ tends to drive the time-evolved wave-function along the chain of Krylov basis vectors with an exponentially increasing average position.\\ \\
To unpack this, it is instructive to consider the derivative of the return amplitude,
\begin{equation}
    R'(t) =  (R(t)^3)\left( \frac{(1 + c^2)(k - \Omega^2)^2 }{8 \sqrt{k} \Omega^2} \sin(2 \sqrt{k} t) -\frac{1-c^2}{2} \gamma \sinh(2 \gamma t) - i c \gamma \cosh(2 \gamma t)\right)\,.
\end{equation}
When $\gamma \rightarrow 0$, the derivative vanishes an infinite number of times, in line with the periodicity of the return amplitude.  When $\gamma$ is non-zero but small it is possible for the derivative to vanish a finite number of times.  This corresponds, of course, to local maxima of the overlap of the time-evolved state with the zero complexity state, $|K_0\rangle$.  The competition between the trigonometric functions (that drive elliptic orbits) and the hyperbolic functions (that drive hyperbolic orbits) gives rise to a rate of growth that is slower than exponential at early times, see Fig. (\ref{zeroCouplingPlots}).   \\ \\
\noindent
An important point in this argument is that it is only applicable in the underdamped ($\varkappa > 0$) regime; in the overdamped regime, the return amplitude is constituted only of hyperbolic functions. In the underdamped regime the Hamiltonian $H_0$ generates elliptic orbits on the manifold of coherent states (see appendix B). In contrast the Hamiltonian $H_1$ always generates hyperbolic orbits.

\subsection{Limit of $\gamma \rightarrow 0$ }

With the damping term taken to zero, the Hamiltonian for the coupled system becomes $H_0 + H_{int}$ and the underlying algebra comprises of two copies of $su(1,1)$, spanned by the operators\footnote{For concreteness, we focus on the underdamped regime.}  
\begin{equation}
    \left\{ \frac{1}{2}A^2, \frac{1}{2}(A^\dag)^2, \frac{1}{4}\left(  A^\dag A + A A^\dag \right) \right\} \cup \left\{ \frac{1}{2}B^2, \frac{1}{2}(B^\dag)^2, \frac{1}{4}\left(  B^\dag B + B B^\dag \right) \right\} \,.
\end{equation}
For a single oscillator we can compute the spread complexity analytically. However, for two oscillators, the combinatoric problem of computing the Krylov basis becomes quite nontrivial. For a single oscillator, say
\begin{equation}
H' = \frac{\beta}{4} \left( A^\dag A + A A^\dag  \right) + \frac{\alpha}{2}(A^\dag A^\dag + A A)
\end{equation}
and an arbitrary coherent-state reference state
\begin{equation}
|z_0\rangle = N e^{\frac{z_0}{2}A^\dag A^\dag } |0, 0\rangle_{AB}
\end{equation}
we can compute
\begin{eqnarray}
& & \langle \bar{z}_0 | e^{- i t H'} |z_0\rangle  \nonumber \\
&=&  \left( \cosh\left(t \sqrt{\alpha^2 - \frac{\beta^2}{4}} \right) + i \frac{2 (z_0 + \bar{z}_0) \alpha + (1 + z_0 \bar{z}_0) \beta   }{(1 - z_0 \bar{z}_0)\sqrt{4 \alpha^2 - \beta^2}   }\sinh\left(t \sqrt{\alpha^2 - \frac{\beta^2}{4}} \right)  \right)^{-\frac{1}{2}} \,.   \nonumber
\end{eqnarray}
This is precisely of the form (\ref{AnalyticRtForm}) so that the Lanczos coefficients and probability amplitudes are known in full detail. The judicious use of some elementary trigonometric identities puts this into the form
\begin{equation}
\left( \frac{\sin( i \omega_1 t + \phi_1
 )}{\sin \phi_1}    \right)^h \,.
\end{equation}
Coupling the oscillators tensors their respective Hilbert spaces and results in a return amplitude that is a product of factors of the above type i.e.
\begin{eqnarray}
& & \left( \frac{\sin( i \omega_1 t + \phi_1
 )}{\sin \phi_1}    \right)^h \left( \frac{\sin( i \omega_2 t + \phi_2
 )}{\sin \phi_2}    \right)^h   \nonumber \\
 & = & \left( \frac{ \sin^2(i \frac{\omega_1 + \omega_2}{2}t + \frac{\phi_1 + \phi_2}{2})   -   \sin^2(i \frac{\omega_1 - \omega_2}{2}t + \frac{\phi_1 - \phi_2}{2}) 
  }{\sin^2\left( \frac{\phi_1 + \phi_2}{2}   \right) -\sin^2\left( \frac{\phi_1 - \phi_2}{2}   \right) }    \right)^h \,.
\end{eqnarray}
If the frequencies $\omega_1, \omega_2$ and factors $\phi_1, \phi_2$ have a small difference then it is clear that the tensor product is approximately of the form for the overlap of a single oscillator but with $h \rightarrow 2h$.  
\subsection{General Case}

The return amplitude can in fact be computed in full generality and for an arbitrary choice of coherent state as the reference state, using the BCH formulae detailed in appendix A. This is a consequence of the Hamiltonian being an element of the finite-dimensional algebra discussed in section 2. The general coherent states are parameterised by three complex parameters.  The general return amplitude is then a sum of exponentials of two frequencies raised to a power
\begin{equation}
    \left( 1 + c_{++} e^{(\omega_1 + \omega_2) t} +   c_{+-} e^{(\omega_1 - \omega_2) t} + c_{-+} e^{(-\omega_1 + \omega_2) t} + c_{--} e^{-(\omega_1 + \omega_2) t}  \right)^{-1/2} \,,
\end{equation}
where 
\begin{equation}
i \omega_1  =  \sqrt{\varkappa - \gamma^2  - \sqrt{\epsilon^2 - 4 \gamma^2 \varkappa}}\,, \ \ \ 
i \omega_2  =  \sqrt{\varkappa - \gamma^2 + \sqrt{\epsilon^2 - 4 \gamma^2 \varkappa}} \,.
   \label{frequencies} 
\end{equation}
The qualitative features of the two-point function (and the resulting spread complexity) are determined by whether these frequencies are real, complex or purely imaginary.  The complex coherent state parameters only affect the value of the $c_{\pm \pm}$ coefficients above. As such, little insight is to be gained from a general choice of coherent state reference state and without loss of generality we restrict to the subset
\begin{equation}
\langle 0, 0 |e^{ \frac{i}{4} \log(\Omega) (x p_x + p_x x + y p_y + p_y y)   }    e^{- i t H} e^{ -\frac{i}{4} \log(\Omega) (x p_x + p_x x + y p_y + p_y y)   }  |0, 0\rangle = \left(   \frac{\Sigma^2 - \Delta^2 - \tau^2 }{16( \omega_2^2 - \omega_1^2  ) \Omega^2 }   \right)^{-\frac{1}{2}}\,,
\end{equation}
where 
\begin{eqnarray}
\Sigma & = & -2 \Omega(\omega_1^2 - \omega_2^2)\left( \cosh(\omega_1 t) + \frac{i \epsilon}{2 \Omega \omega_1} \sinh(\omega_1 t) +   \cosh(\omega_2 t) + \frac{i \epsilon}{2 \Omega \omega_2} \sinh(\omega_2 t)  \right)     \nonumber \\
 & & + 2 i \epsilon (\varkappa + \Omega^2)\left( \frac{ 
  \sinh(\omega_1 t)}{\omega_1} -  \frac{\sinh(\omega_2 t)}{\omega_2}    \right)\,,     \nonumber \\
  \Delta & = & 4 \epsilon \Omega \left( \cosh(\omega_2 t) + \frac{i \epsilon}{2 \Omega} \frac{\sinh(\omega_2 t)}{\omega_2} - \cosh(\omega_1 t) - \frac{i \epsilon}{2 \Omega} \frac{\sinh(\omega_1 t)}{\omega_1}    \right) \nonumber \\
   & & + i(\varkappa + \Omega^2)\left( (4 \gamma^2 + \omega_1^2 - \omega_2^2) \frac{\sinh(\omega_1 t)}{\omega_1} + (-4 \gamma^2 + \omega_1^2 - \omega_2^2) \frac{\sinh(\omega_2 t)}{\omega_2}  \right)\,,     \nonumber \\
   \tau & = & 4\gamma\left( (\varkappa-\Omega^2)(\cosh(\omega_2 t) - \cosh(\omega_1 t)) + i \epsilon \Omega\left( 
    \frac{\sinh(\omega_1 t) }{\omega_1} -  \frac{\sinh(\omega_2 t) }{\omega_2}  \right)    \right)\,.
\end{eqnarray}
 The scaling of the oscillators annihilating the reference state, $\Omega$, clearly does not affect the time-dependence of the trigonometric functions, only the coefficients in the superposition. By tuning $\Delta$ we are able to dial the values of the coefficients multiplying the exponentials with frequencies $\omega_1$ and $\omega_2$. \\ \\
To determine the spread complexity, we will make use of the Lanczos algorithm (\ref{LanczosProb}) to first compute the probability amplitudes.  To accurately represent the time-evolved state up to a desired time we need to include a sufficient number of probability amplitudes.

\subsection{Underdamped Regime}

\begin{figure}
\centering
\includegraphics[width=0.6\textwidth]{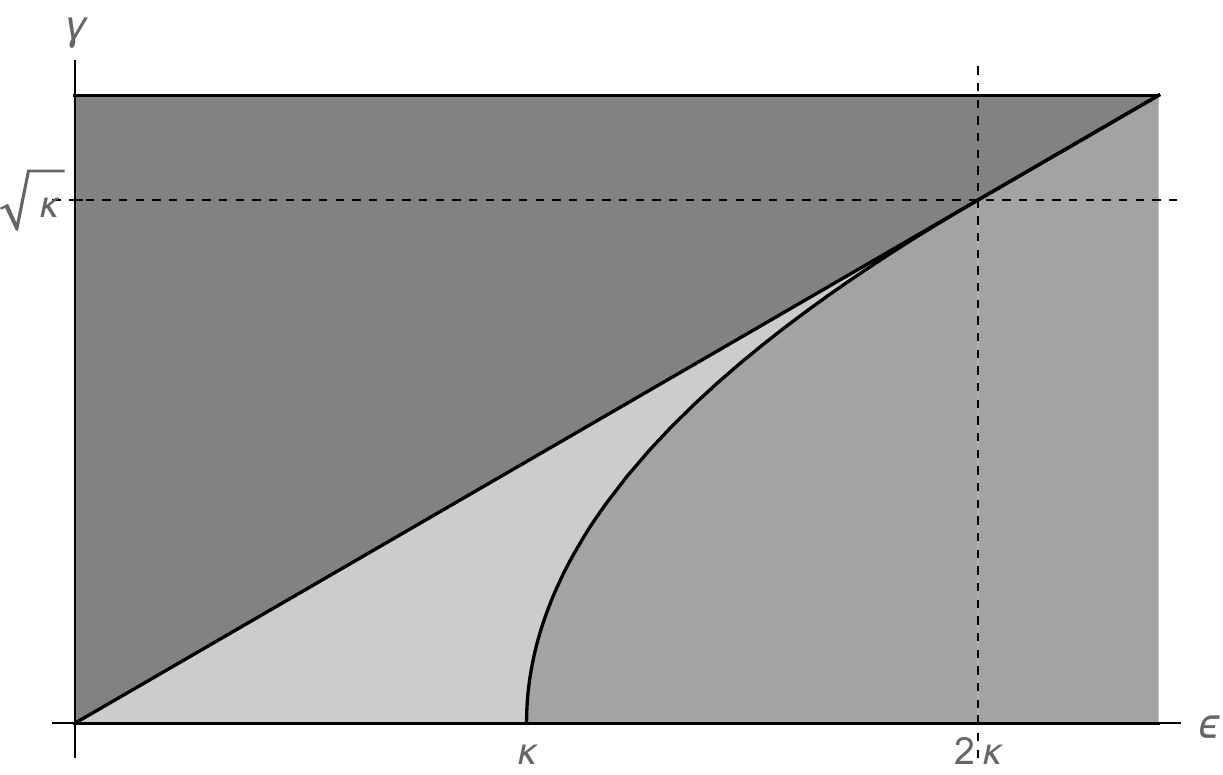}
\caption{In the underdamped case there are three distinct regimes for the frequencies. If $\epsilon > 2 \sqrt{\varkappa}\gamma$ and $\epsilon < \gamma^2 + \varkappa$ the frequencies are real - this corresponds to the lightest shading above. This window closes as $\gamma$ approaches $\sqrt{\varkappa}$ and after $\gamma > 2\sqrt{\varkappa}$ it is no longer possible to obtain two real frequencies   }  
\label{underDampedPhases}
\end{figure}

We first focus on the underdamped region of phase space, characterised by $\varkappa > 0$. In this regime there are three distinct regions for the frequencies \cite{Bender:2013qta}, as sketched in Fig. (\ref{underDampedPhases}). 
We find that the simplest expressions are obtained when $\Omega = \sqrt{\varkappa}$ and so restrict to this value below.
\subsubsection*{Weak Coupling}

In the region $\epsilon < \epsilon_1$, the system is weakly coupled and the return amplitude,
\begin{eqnarray}
\langle 0, 0 | e^{- i t H} |0,0\rangle & = & \left( \frac{\Omega^2 \gamma^2 
 }{2 \Delta^2 \Sigma^2} - \frac{(\Omega^2 - \Sigma^2)( (\Omega^2 - \Sigma^2) \Delta^2 + \Omega^4  ) 
  }{2 \Sigma^2 (\Delta^2 + \Sigma^2) \Omega^2} \cos(2 \Sigma t)   \right. \nonumber \\
  && + \frac{(\Sigma^2 -\Omega^2)^{\frac{3}{2}} \sqrt{\Delta^2 + \Omega^2} }{\Sigma (\Delta^2 + \Sigma^2) \Omega} \sin(2 \Sigma t) + \frac{(\Sigma^2 -\Omega^2)^{\frac{1}{2}} (\Delta^2 + \Omega^2)^{\frac{3}{2} } }{\Delta (\Delta^2 + \Sigma^2) \Omega} \sinh(2 \Delta t)    \nonumber \\
  & & \left.  + \frac{ (\Delta^2 + \Omega^2)( \Delta^2 \Sigma^2 + \Omega^2 (\Sigma^2 - \Omega^2)  )    }{2 \Delta^2 (\Delta^2 + \Sigma^2) \Omega^2} \cosh(2 \Delta t)   \right)^{-\frac{1}{2}}\,,
\end{eqnarray}
where now
\begin{eqnarray}
2\Sigma & = & \sqrt{\Omega^2 - \gamma^2 + i \sqrt{4 \gamma^2 \Omega^2 - \epsilon^2}  } + \sqrt{\Omega^2 - \gamma^2 - i \sqrt{4 \gamma^2 \Omega^2 - \epsilon^2}  }   \nonumber \\
2\Delta & = & i\left( \sqrt{\Omega^2 - \gamma^2 + i \sqrt{4 \gamma^2 \Omega^2 - \epsilon^2}  } - \sqrt{\Omega^2 - \gamma^2 - i \sqrt{4 \gamma^2 \Omega^2 - \epsilon^2}  }    \right)\,.    \label{DeltaSigmaDef}
\end{eqnarray} 
In this regime the return amplitude is made up of hyperbolic functions with time-scale $\Delta$ and trigonometric functions with time-scale $\Sigma$. When $\Delta$ is small and the coefficients multiplying the trigonometric functions are comparable to those multiplying the hyperbolic ones the spread complexity is oscillatory. This is quite apparent in the early time regime but is also evident even at late times. The presence of the hyperbolic functions immediately implies that the probability amplitudes will decay so that they are not localised on the Krylov chain. Consequently, the spread complexity has an overall growing profile and keeps on growing indefinitely. 
  \begin{figure}[h]
  \centering
\includegraphics[trim={0 0 0 1cm},clip,width = 0.6\textwidth]{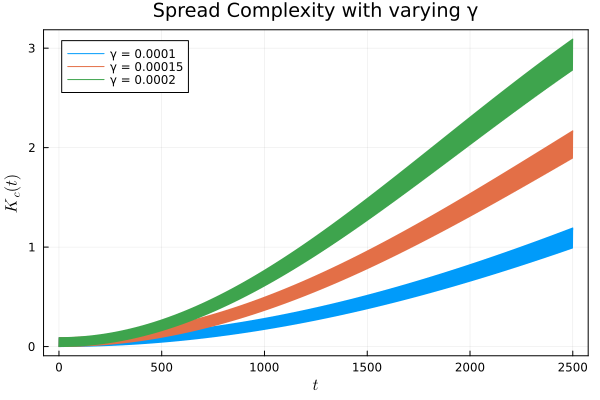}
\caption{The spread complexity for $\varkappa = 1.5,\ \Omega = 1.0 ,\ N = 30, \text{ and } \epsilon = 0.5 \epsilon_{1}$ for various values of $\gamma$. At least, 95\% of the probability is captured. The exponential growth is more pronounced for larger values of $\gamma$. The parameters are sufficiently small that choosing $\epsilon = 0$ does not affect the complexity, since the system is very weakly coupled. The "band-like" appearance is indicating closely-spaced oscillations.}
\label{epsilon0p1Plot}
 \end{figure}

This is sketched in Fig. (\ref{epsilon0p1Plot}). The spread complexity grows exponentially, with time-scale that is approximated by the quantity $\Delta$ in (\ref{DeltaSigmaDef}). In the weak-coupling regime the probability amplitudes also contain trigonometric functions, with frequency $\Sigma$. If $\Sigma$ is large compared to $\Delta$ we obtain an overall exponentially growing profile, with oscillations. An extreme example of this has already been unpacked in Fig. (\ref{zeroCouplingPlots}).  

\subsubsection*{Rabi Oscillations}

\begin{figure}[h!]
  \centering
\includegraphics[trim={0 0 0 1cm},clip,width = 0.6\textwidth]{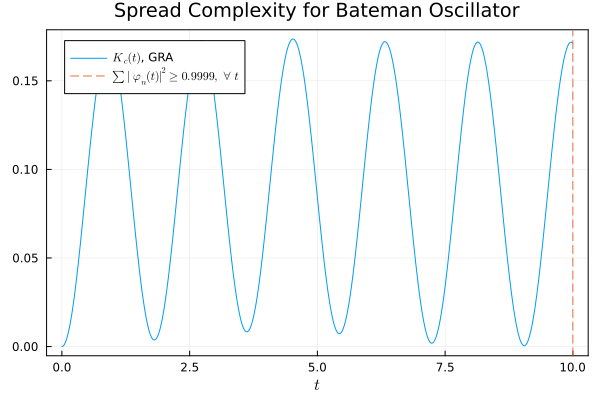}
\caption{The spread complexity approximated with the first $40$ probability amplitudes for $\varkappa = 2.0,\ \Omega =1.0$, $\gamma = 0.01$, $\epsilon = 1.014$.  At least 99.99\% of the time-evolved reference state is thus captured by the first 40 complexity wavefunctions for the displayed time-range.}
\label{spreadPlotRabi}
 \end{figure}

The $\mathsf{PT}$-symmetric phase where the coupled system exhibits Rabi oscillations is obtained when 
$\epsilon_1 < \epsilon < \epsilon_2$. Here we find the following expression for the return amplitude
\begin{eqnarray}
\langle 0, 0 | e^{- i t H} |0,0\rangle & = & \left( -\frac{\Omega^2 \gamma^2 
 }{2 \Delta^2 \Sigma^2} + \frac{(\Omega^2 - \Sigma^2)( -(\Omega^2 - \Sigma^2) \Delta^2 + \Omega^4  ) 
  }{2 \Sigma^2 (\Delta^2 + \Sigma^2) \Omega^2} \cos(2 \Sigma t)   \right. \nonumber \\
  && - i\frac{(\Sigma^2 -\Omega^2)^{\frac{3}{2}} \sqrt{\Delta^2 - \Omega^2} }{\Sigma ( \Sigma^2 - \Delta^2) \Omega} \sin(2 \Sigma t) + i \frac{(\Sigma^2 -\Omega^2)^{\frac{1}{2}} (\Delta^2 - \Omega^2)^{\frac{3}{2} } }{\Delta (\Delta^2 - \Sigma^2) \Omega} \sin(2 \Delta t)    \nonumber \\
  & & \left.  + \frac{ (\Delta^2 - \Omega^2)( \Delta^2 \Sigma^2 + \Omega^4 - \Sigma^2 \Omega^2    }{2 \Delta^2 (\Delta^2 - \Sigma^2) \Omega^2} \cos(2 \Delta t)   \right)^{-\frac{1}{2}}\,,
\end{eqnarray}
where now
\begin{eqnarray}
2\Sigma & = & \sqrt{\Omega^2 - \gamma^2 +  \sqrt{ \epsilon^2 - 4 \gamma^2 \Omega^2}  } + \sqrt{\Omega^2 - \gamma^2 -  \sqrt{\epsilon^2  - 4 \gamma^2 \Omega^2  } }  \nonumber \\
2\Delta & = & \sqrt{\Omega^2 - \gamma^2 + \sqrt{\epsilon^2 - 4 \gamma^2 \Omega^2}  } - \sqrt{\Omega^2 - \gamma^2 - \sqrt{\epsilon^2 - 4 \gamma^2 \Omega^2}  }  \,.
\end{eqnarray}
In this region the return amplitude (and, by extension, the spread complexity) consists completely of periodic functions, which is consistent with the observation that this region supports all real frequencies \cite{Bender:2013qta} and the Hamiltonian possesses a discrete spectrum. The probability amplitudes are also expressed in terms of these periodic functions. For rational $\omega_1, \omega_2$ it is easy to show that the system is periodic, repeating whenever $\Sigma t$ and $\Delta t$ are simultaneously multiples of $\pi$. In general, the real numbers can be approximated by rational ones, bringing the system within some tolerance of being periodic.  Crucially, this implies that spread complexity is \textit{bounded} in the phase of Rabi oscillations. This is a special feature that is only possible and always true in this special regime of parameter space. An example is plotted in Fig. (\ref{spreadPlotRabi})

 \subsubsection*{Ultra-Strong Coupling}
 
\begin{figure}[h!]
\centering
\includegraphics[trim={0 0 0 1cm},clip,width = 0.6\textwidth]{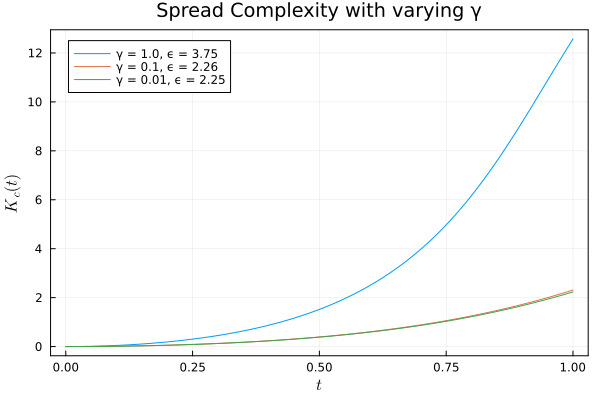}
\caption{The spread complexity for $\varkappa = 1.5,\ \Omega =1$, and $N = 80$ for various values of $\gamma$. At least, 99\% of the probability is captured. The exponential growth is more pronounced for larger values of $\gamma$ due to a corresponding increment in $\epsilon$, fixed at $1.5 \epsilon_{2}$ for each of the three cases, since the system is strongly coupled.}
\label{spreadPlotStrongInt}
\end{figure}

Finally, in the much less studied ultra-strong interaction region $\epsilon > \epsilon_2$ we find that 
\begin{eqnarray}
& & \langle 0, 0| e^{-i t H} |0, 0\rangle   \nonumber \\
& = & \left( -\frac{\Omega^2 (\Omega^2 + 2 (\omega_1^2 - \omega_2^2))}{2(\omega_1^2 + \omega_2^2)^2}   + 
  \frac{ \Omega^4 + 2 \Omega^2(\omega_1^2 - \omega_2^2) + 2 (\omega_1^2 + \omega_2^2)^2  }{2(\omega_1^2 + \omega-2^2)^2} \cosh(2 \omega_1 t) \cos(2 \omega_2 t)    \right.  \nonumber \\
& &+ i \frac{\sqrt{\Omega^4 + 2 \Omega^2(\omega_1^2 - \omega_2^2) + (\omega_1^2 + \omega_2^2)^2 }}{2 \Omega (\omega_1^2 + \omega_2^2)}
\times\\ 
& &\times\left( \frac{\Omega^2 + \omega_1^2 + \omega_2^2}{\omega_2}  \cosh(2 \omega_1 t) \sin(2\omega_2 t) + \frac{\Omega^2 - \omega_1^2 - \omega_2^2}{\omega_1}   \sinh(2 \omega_1 t) \cos(2 \omega_2 t) \right)    \nonumber \\
&  & \left. -\frac{\Omega^6 (\omega_1^2 - \omega_2^2) + 2 \Omega^4 (\omega_1^2 -  \omega_2^2)^2 + 2 \Omega^2(\omega_1^2 - \omega_2^2)(\omega_1^2 + \omega_2^2)^2 + (\omega_1^2 + \omega_2^2)^4}{4 \Omega^2 \omega_1 \omega_2 (\omega_1^2 + \omega_2^2)^2} \sinh(2 \omega_1 t) \sin(2 \omega_2 t)    \right)^{-\frac{1}{2}  }\nonumber
\end{eqnarray}
 where 
 \begin{equation}
 \omega_1  =  \frac{1}{2} \sqrt{ \sqrt{\epsilon^2 - 4 \gamma^2 \Omega^2} - (\Omega^2 - \gamma^2)  }\,,  \ \ \ \ 
 \omega_2  =  \frac{1}{2} \sqrt{ \sqrt{\epsilon^2 - 4 \gamma^2 \Omega^2} + (\Omega^2 - \gamma^2)  }\,.
 \end{equation}
Here, every term in the return amplitude grows exponentially and we see no periodic or oscillatory regimes. \\ \\ 
In the ultra-strong interaction region, we again find that the spread complexity grows exponentially; see Fig. (\ref{spreadPlotStrongInt}). We also note that, in order to capture the time-evolved reference state up to a desired time, one requires more probability amplitudes than in the weak interaction regime.  

\subsection{Overdamped Regime}

In the overdamped regime ($\varkappa <0$) the frequencies (\ref{frequencies}) cannot be tuned to be real numbers simultaneously.  The simplest expression for the return amplitude is obtained when $\varkappa = -\Omega^2$ which takes the surprisingly simple form
\begin{equation}
R(t) = \left(\left(  \cosh(\omega_1 t) + \frac{i \epsilon}{2 \sqrt{-\varkappa} \omega_1}  \sinh(\omega_1 t)   \right)\left(  \cosh(\omega_2 t) + \frac{i \epsilon}{2 \sqrt{-\varkappa} \omega_2} \sinh(\omega_2 t)    \right)  \right)^{-\frac{1}{2}}\,.
\end{equation}
The frequencies here are
\begin{equation}
\omega_1 =  \sqrt{ \Omega^2 + \gamma^2 - \sqrt{\epsilon^2 + 4 \Omega^2 \gamma^2}     }\,,\ \ 
\omega_2  =  \sqrt{ \Omega^2 + \gamma^2 + \sqrt{\epsilon^2 + 4 \Omega^2 \gamma^2}     }   \nonumber
\end{equation}
and $\omega_1$ becomes pure imaginary when $\epsilon > | \gamma^2 + \varkappa| = \epsilon_2$. There are thus two distinct phases; one where the return amplitude consists only of hyperbolic functions and the other with a combination of trigonometric and hyperbolic functions. The former is formally similar to the ultra-strongly coupled, underdamped oscillator while the latter to the weakly-coupled underdamped oscillator. Unlike the underdamped oscillator, however, we always have $\omega_1 < \omega_2$. This has important consequences since the hyperbolic functions grow faster than the trigonometric functions.  As such, we cannot find parameter regimes where drivers of elliptic orbits have a significant impact on the growth of complexity as compared to the drivers of hyperbolic orbits. Spread complexity thus appears to (essentially) grow exponentially in the overdamped regime.

 \begin{figure}[ h ]
	\centering

    \begin{subfigure}{0.45\linewidth}
        \includegraphics[trim={0 0 0 1cm},clip,width = \linewidth]{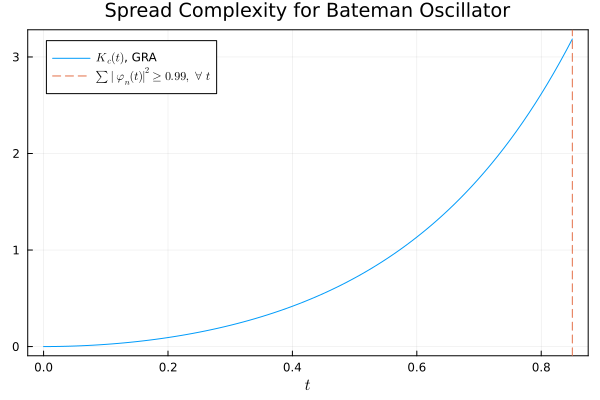}
	   \caption{${ \epsilon = 0.007 \epsilon_{2} }$}
	   \label{fig:ComplexityOverDampedWeak}
    \end{subfigure}
    \hfill
    \begin{subfigure}{0.45\linewidth}
        \includegraphics[trim={0 0 0 1cm},clip,width = \linewidth]{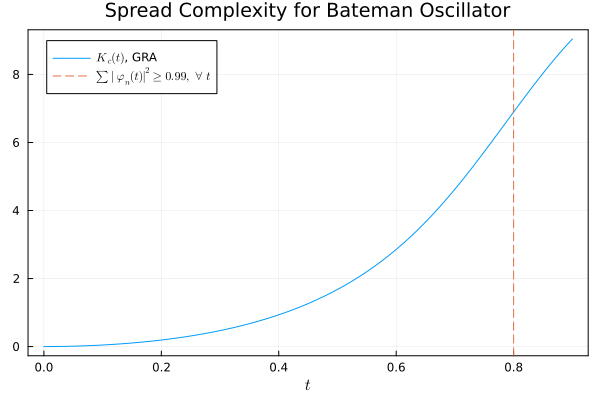}
	   \caption{${ \epsilon = 1.5 \epsilon_{2} }$}
	   \label{fig:ComplexityOverDampedStrong}
    \end{subfigure}

    \caption{The spread complexity for over-damped oscillator system with ${ \varkappa=-2.0, \ \epsilon_{2} = 2.0,\ \gamma = 0.01,\ \Omega = 1.0}$, and, ${ N = 50 }$.}
    \label{fig:ComplexityCriticallyDamped}

\end{figure}

 \subsection{Critically Damped Regime}

 Finally, in the critically damped regime ($\varkappa = 0$) we find the return amplitude,
 \begin{eqnarray}
&& R(t) \nonumber \\
&& = \left( \left( 
  \cosh(\sqrt{\gamma^2 + \epsilon} t) + i \frac{(\epsilon - \Omega^2) \sinh(\sqrt{\gamma^2 + \epsilon} t)}{2\Omega \sqrt{\gamma^2 + \epsilon}}    \right) \left( 
  \cosh(\sqrt{\gamma^2 - \epsilon} t) + i \frac{(\epsilon + \Omega^2)\sinh(\sqrt{\gamma^2 - \epsilon} t)}{2\Omega \sqrt{\gamma^2 + \epsilon}}    \right)    \right.   \nonumber \\
  & & + \left. 
 \frac{\gamma^2 \Omega^2}{2 \epsilon^2}\left(  \cosh(\sqrt{\gamma^2 - \epsilon} t) \cosh(\sqrt{\gamma^2 + \epsilon} t) - \frac{\gamma^2 \sinh(\sqrt{\gamma^2 - \epsilon} t) \sinh(\sqrt{\gamma^2 + \epsilon} t)}{\sqrt{\gamma^4 - \epsilon^2}   } -1 \right)    \right)^{-\frac{1}{2}} \,.
 \end{eqnarray}
 In this case there is a single transition point at $\epsilon = \gamma^2$.  Again, this is characterised by the return amplitude containing only hyperbolic trigonometric functions of both hyperbolic functions and trigonometric functions.  Similar to the overdamped regime, the time-scale of the drivers of hyperbolic orbits is always smaller than the drivers of elliptic orbits and exponential growth always dominates.

 \begin{figure}[ h ]
	\centering

    \begin{subfigure}{0.45\linewidth}
        \includegraphics[trim={0 0 0 1cm},clip,width = \linewidth]{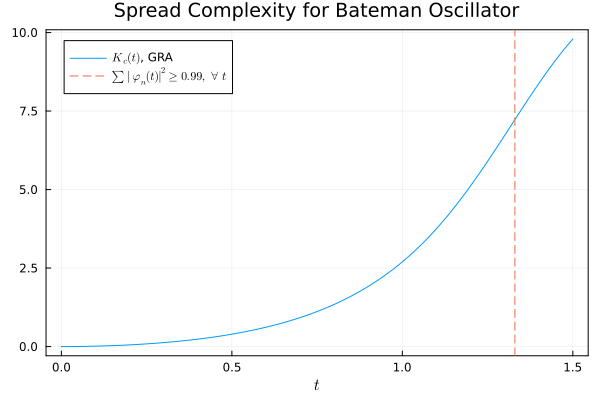}
	   \caption{${ \epsilon = 0.5 \times \epsilon_{2} }$}
	   \label{fig:ComplexityCriticallyDampedWeak}
    \end{subfigure}
    \hfill
    \begin{subfigure}{0.45\linewidth}
        \includegraphics[trim={0 0 0 1cm},clip,width = \linewidth]{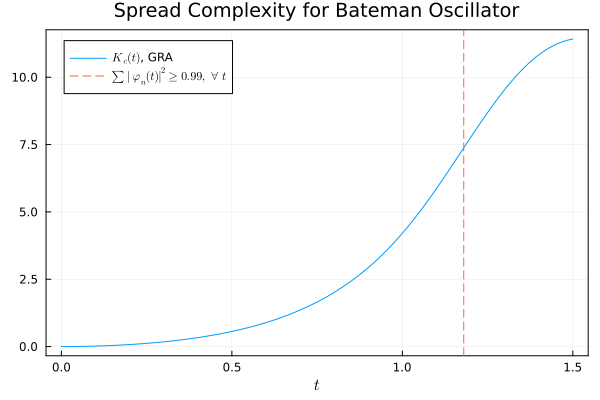}
	   \caption{${ \epsilon = 1.5 \times \epsilon_{2} }$}
	   \label{fig:ComplexityCriticallyDampedStrong}
    \end{subfigure}

    \caption{The spread complexity for critically-damped Bateman oscillator system with ${ \varkappa = 0.0,\ \epsilon_{2} = 1.0,\ \gamma = 1.0,\ \Omega = 1.0}$, and, ${ N = 50 }$.}
    \label{fig:ComplexityCriticallyDamped}

\end{figure}

\section{Conclusions}

It is, by now, well established that spread complexity encodes information about the phase of quantum matter. Two notable examples include the chaos/integrable transition in the Ising and SYK models and topological phase transitions in 1-dimensional p-wave superconductors. In this article, we have investigated whether the spread complexity can also detect $\mathsf{PT}$-symmetric transitions, which occur when a system with a balanced gain and loss of energy undergoes spontaneous symmetry breaking. We have focused on the Bateman oscillator, which exhibits a $\mathsf{PT}$-symmetric transition when coupled to a quadratic potential with a large enough strength. We have computed the spread complexity from the return amplitude, which in turn is directly related to the spectrum of the Hamiltonian. Our results show that the spread complexity is indeed sensitive to the $\mathsf{PT}$-symmetric transition, which manifests as the emergence or disappearance of a well-defined ground state.  \\ \\
More surprising, however, is that spread complexity is also able to distinguish between the ultra-strong coupling and weak coupling regimes, both of which fall outside the $\mathsf{PT}$-symmetric phase.  The key feature that makes this possible is that spread complexity depends on both frequencies characterizing the families of symmetry generators.  In the $\mathsf{PT}$ symmetric phase these frequencies are both real so that spread complexity is a function of trigonometric functions.  In the weak coupling regime the frequencies are hermitian conjugates of each other, while there is one real and one purely imaginary frequency in the strong coupling regime.  Consequently, the spread complexity generated by a Hamiltonian in the ultra-strong coupling regime exhibits exponential growth at all times, while one from the weak coupling regime can (in the underdamped phase) have much slower growth at early times.  This statement is not universal and depends on the specifics of the chosen reference state; some choices of reference state are blind to this difference.  This should be contrasted with the $\mathsf{PT}$ symmetric phase transition which is robust and universal in the sense that it can be identified for any choice of reference state.  \\ \\
The Bateman oscillator, as a set of two coupled bosonic oscillators, acts on an infinite dimensional Hilbert space.  As such, the Krylov space is infinite dimensional and the complexity may be unbounded.  Indeed, as our results confirm, complexity outside of the $\mathsf{PT}$ symmetric phase grows exponentially at late times and continues to do so indefinitely.  This is closely tied to the absence of normalisable eigenstates so that the return amplitude cannot be expanded in such a basis.  This is, however, possible in the $\mathsf{PT}$ symmetric phase as is reflected by the periodic functions that make up the Krylov probability amplitudes.  In this phase more of the usual intuition for complexity growth (and saturation) applicable to finite dimensional Krylov spaces holds.  As a corollary to these results it is possible to eliminate candidate coherent state eigenstates of the system by selecting them as reference states; if the complexity is non-trivial then the reference state cannot be an eigenstate of the system.  \\ \\
We have stressed in the main body of the text that these features of spread complexity can largely be understood from the frequencies of the system. These, in turn, are intimately related to the combination of elliptic, parabolic and hyperbolic orbits generated by the Hamiltonian on the manifold of coherent states. However, one subtlety to bear in mind is that the complexity of a coherent state target state depends sensitively on the choice of Hamiltonian, unlike the unit-rank cases discussed in Appendix B.  Nevertheless, the frequencies do end up encoding a substantial amount of information about the behavior of spread complexity, and we suspect that this is a general feature for both bosonic and fermionic systems.  Making this link precise and classifying the coherent state reference states along these lines for more general Hamiltonians would be an interesting future direction to explore, particularly in open quantum systems with mixed system/bath statistics. \\ \\
The Bateman system consists of one oscillator which is damped and another which is anti-damped, controlled by the parameter $\gamma$, chosen so that the energy loss and gain is exactly balanced in the combined system. As we have seen in this very simple system, this dynamical equilibrium phase offers a remarkable simplification of the dynamics. Cycling back to the opening question in the Introduction, it would be of tremendous interest to perturb the system away from equilibrium and compare the rate of complexity growth in these regimes as the system equilibrates.  \\ \\
Finally, it is worth noting that the algebra underlying the Bateman oscillator is a real form of $so(5, \mathbb{C})$.  As such the analysis we have performed here is closely related to $so(3,2)$, the Euclidean conformal group in three dimensions.  Understanding the dependence of spread complexity on the scaling dimension and spin of the reference state would be an important step in clarifying the holographic dual description of spread complexity.

\acknowledgments

We would like to thank Javier Magan, Horatiu Nastase, Francesco Petruccione and especially Dario Rosa for insightful discussions. JM and NG would like to acknowledge support from the ICTP through the Associates Programme, from the Simons Foundation (Grant No. 284558FY19). SSH, JM and HJRVZ are supported in part by the ``Quantum Technologies for Sustainable Development''  grant from the National Institute for Theoretical and Computational Sciences of South Africa (NITHECS). NG is supported by a Faculty of Science PhD Fellowship at the University of Cape Town. CB would like to thank the Isaac Newton Institute for Mathematical Sciences for support and hospitality during the programme \textit{Black holes: bridges between number theory and holographic quantum information}, where the majority of this work was completed and supported by: EPSRC grant number EP/R014604/1. CB is supported by the Oppenheimer Memorial Trust PhD Fellowship and the Harry Crossley PhD Fellowship.


\bibliographystyle{JHEP}
\bibliography{BatemanRefs}

\providecommand{\href}[2]{#2}\begingroup\raggedright\begin{thebibliography}{10}

\bibitem{Sachdev:2011fcc}
S.~Sachdev, \emph{{Quantum Phase Transitions}}, Cambridge University Press (4,
  2011),
  \href{https://doi.org/10.1017/cbo9780511973765}{10.1017/cbo9780511973765}.

\bibitem{Eisert:2014jea}
J.~Eisert, M.~Friesdorf and C.~Gogolin, \emph{{Quantum many-body systems out of
  equilibrium}}, \href{https://doi.org/10.1038/nphys3215}{\emph{Nature Phys.}
  {\bfseries 11} (2015) 124} [\href{https://arxiv.org/abs/1408.5148}{{\ttfamily
  1408.5148}}].

\bibitem{Heyl:2018jzi}
M.~Heyl, \emph{{Dynamical quantum phase transitions: a brief survey}},
  \href{https://doi.org/10.1209/0295-5075/125/26001}{\emph{EPL} {\bfseries 125}
  (2019) 26001} [\href{https://arxiv.org/abs/1811.02575}{{\ttfamily
  1811.02575}}].

\bibitem{Wald:1997qz}
R.M.~Wald, \emph{{Black holes and thermodynamics}},  in \emph{{Symposium on
  Black Holes and Relativistic Stars (dedicated to memory of S.
  Chandrasekhar)}}, 2, 1997
  [\href{https://arxiv.org/abs/gr-qc/9702022}{{\ttfamily gr-qc/9702022}}].

\bibitem{Wald:1999vt}
R.M.~Wald, \emph{{The thermodynamics of black holes}},
  \href{https://doi.org/10.12942/lrr-2001-6}{\emph{Living Rev. Rel.} {\bfseries
  4} (2001) 6} [\href{https://arxiv.org/abs/gr-qc/9912119}{{\ttfamily
  gr-qc/9912119}}].

\bibitem{Barbon:2003aq}
J.L.F.~Barbon and E.~Rabinovici, \emph{{Very long time scales and black hole
  thermal equilibrium}},
  \href{https://doi.org/10.1088/1126-6708/2003/11/047}{\emph{JHEP} {\bfseries
  11} (2003) 047} [\href{https://arxiv.org/abs/hep-th/0308063}{{\ttfamily
  hep-th/0308063}}].

\bibitem{PhysRevLett.63.1070}
K.Y.~Min, J.~Stavans, R.~Piazza and W.I.~Goldburg, \emph{Steady-state
  nucleation in a binary mixture: The effect of stirring},
  \href{https://doi.org/10.1103/PhysRevLett.63.1070}{\emph{Phys. Rev. Lett.}
  {\bfseries 63} (1989) 1070}.

\bibitem{Haque:2021kdm}
S.S.~Haque, C.~Jana and B.~Underwood, \emph{{Saturation of thermal complexity
  of purification}}, \href{https://doi.org/10.1007/JHEP01(2022)159}{\emph{JHEP}
  {\bfseries 01} (2022) 159}
  [\href{https://arxiv.org/abs/2107.08969}{{\ttfamily 2107.08969}}].

\bibitem{Bhattacharyya:2021fii}
A.~Bhattacharyya, T.~Hanif, S.S.~Haque and M.K.~Rahman, \emph{{Complexity for
  an open quantum system}},
  \href{https://doi.org/10.1103/PhysRevD.105.046011}{\emph{Phys. Rev. D}
  {\bfseries 105} (2022) 046011}
  [\href{https://arxiv.org/abs/2112.03955}{{\ttfamily 2112.03955}}].

\bibitem{Bhattacharyya:2020iic}
A.~Bhattacharyya, S.S.~Haque and E.H.~Kim, \emph{{Complexity from the reduced
  density matrix: a new diagnostic for chaos}},
  \href{https://doi.org/10.1007/JHEP10(2021)028}{\emph{JHEP} {\bfseries 10}
  (2021) 028} [\href{https://arxiv.org/abs/2011.04705}{{\ttfamily
  2011.04705}}].

\bibitem{Schuster:2022bot}
T.~Schuster and N.Y.~Yao, \emph{{Operator Growth in Open Quantum Systems}},
  \href{https://doi.org/10.1103/PhysRevLett.131.160402}{\emph{Phys. Rev. Lett.}
  {\bfseries 131} (2023) 160402}
  [\href{https://arxiv.org/abs/2208.12272}{{\ttfamily 2208.12272}}].

\bibitem{Bhattacharyya:2022rhm}
A.~Bhattacharyya, T.~Hanif, S.S.~Haque and A.~Paul, \emph{{Decoherence,
  entanglement negativity, and circuit complexity for an open quantum system}},
  \href{https://doi.org/10.1103/PhysRevD.107.106007}{\emph{Phys. Rev. D}
  {\bfseries 107} (2023) 106007}
  [\href{https://arxiv.org/abs/2210.09268}{{\ttfamily 2210.09268}}].

\bibitem{Bhattacharjee:2022lzy}
B.~Bhattacharjee, X.~Cao, P.~Nandy and T.~Pathak, \emph{{Operator growth in
  open quantum systems: lessons from the dissipative SYK}},
  \href{https://doi.org/10.1007/JHEP03(2023)054}{\emph{JHEP} {\bfseries 03}
  (2023) 054} [\href{https://arxiv.org/abs/2212.06180}{{\ttfamily
  2212.06180}}].

\bibitem{Bhattacharya:2022gbz}
A.~Bhattacharya, P.~Nandy, P.P.~Nath and H.~Sahu, \emph{{Operator growth and
  Krylov construction in dissipative open quantum systems}},
  \href{https://doi.org/10.1007/JHEP12(2022)081}{\emph{JHEP} {\bfseries 12}
  (2022) 081} [\href{https://arxiv.org/abs/2207.05347}{{\ttfamily
  2207.05347}}].

\bibitem{Liu:2022god}
C.~Liu, H.~Tang and H.~Zhai, \emph{{Krylov complexity in open quantum
  systems}},
  \href{https://doi.org/10.1103/PhysRevResearch.5.033085}{\emph{Phys. Rev.
  Res.} {\bfseries 5} (2023) 033085}
  [\href{https://arxiv.org/abs/2207.13603}{{\ttfamily 2207.13603}}].

\bibitem{NSSrivatsa:2023qlh}
N.S.~Srivatsa and C.~von Keyserlingk, \emph{{The operator growth hypothesis in
  open quantum systems}},  \href{https://arxiv.org/abs/2310.15376}{{\ttfamily
  2310.15376}}.

\bibitem{Bhattacharya:2023zqt}
A.~Bhattacharya, P.~Nandy, P.P.~Nath and H.~Sahu, \emph{{On Krylov complexity
  in open systems: an approach via bi-Lanczos algorithm}},
  \href{https://doi.org/10.1007/JHEP12(2023)066}{\emph{JHEP} {\bfseries 12}
  (2023) 066} [\href{https://arxiv.org/abs/2303.04175}{{\ttfamily
  2303.04175}}].

\bibitem{Ali:2019zcj}
T.~Ali, A.~Bhattacharyya, S.S.~Haque, E.H.~Kim, N.~Moynihan and J.~Murugan,
  \emph{{Chaos and Complexity in Quantum Mechanics}},
  \href{https://doi.org/10.1103/PhysRevD.101.026021}{\emph{Phys. Rev. D}
  {\bfseries 101} (2020) 026021}
  [\href{https://arxiv.org/abs/1905.13534}{{\ttfamily 1905.13534}}].

\bibitem{Bhattacharyya:2020kgu}
A.~Bhattacharyya, S.~Das, S.S.~Haque and B.~Underwood, \emph{{Rise of
  cosmological complexity: Saturation of growth and chaos}},
  \href{https://doi.org/10.1103/PhysRevResearch.2.033273}{\emph{Phys. Rev.
  Res.} {\bfseries 2} (2020) 033273}
  [\href{https://arxiv.org/abs/2005.10854}{{\ttfamily 2005.10854}}].

\bibitem{Bhattacharyya:2020rpy}
A.~Bhattacharyya, S.~Das, S.~Shajidul~Haque and B.~Underwood,
  \emph{{Cosmological Complexity}},
  \href{https://doi.org/10.1103/PhysRevD.101.106020}{\emph{Phys. Rev. D}
  {\bfseries 101} (2020) 106020}
  [\href{https://arxiv.org/abs/2001.08664}{{\ttfamily 2001.08664}}].

\bibitem{Bhattacharyya:2020art}
A.~Bhattacharyya, W.~Chemissany, S.S.~Haque, J.~Murugan and B.~Yan, \emph{{The
  Multi-faceted Inverted Harmonic Oscillator: Chaos and Complexity}},
  \href{https://doi.org/10.21468/SciPostPhysCore.4.1.002}{\emph{SciPost Phys.
  Core} {\bfseries 4} (2021) 002}
  [\href{https://arxiv.org/abs/2007.01232}{{\ttfamily 2007.01232}}].

\bibitem{Bhattacharyya:2020qtd}
A.~Bhattacharyya, S.S.~Haque, G.~Jafari, J.~Murugan and D.~Rapotu,
  \emph{{Krylov complexity and spectral form factor for noisy random matrix
  models}}, \href{https://doi.org/10.1007/JHEP10(2023)157}{\emph{JHEP}
  {\bfseries 23} (2020) 157}
  [\href{https://arxiv.org/abs/2307.15495}{{\ttfamily 2307.15495}}].

\bibitem{Bhattacharjee:2022vlt}
B.~Bhattacharjee, X.~Cao, P.~Nandy and T.~Pathak, \emph{{Krylov complexity in
  saddle-dominated scrambling}},
  \href{https://doi.org/10.1007/JHEP05(2022)174}{\emph{JHEP} {\bfseries 05}
  (2022) 174} [\href{https://arxiv.org/abs/2203.03534}{{\ttfamily
  2203.03534}}].

\bibitem{Parker:2018yvk}
D.E.~Parker, X.~Cao, A.~Avdoshkin, T.~Scaffidi and E.~Altman, \emph{{A
  Universal Operator Growth Hypothesis}},
  \href{https://doi.org/10.1103/PhysRevX.9.041017}{\emph{Phys. Rev. X}
  {\bfseries 9} (2019) 041017}
  [\href{https://arxiv.org/abs/1812.08657}{{\ttfamily 1812.08657}}].

\bibitem{Magan:2020iac}
J.M.~Mag\'an and J.~Sim\'on, \emph{{On operator growth and emergent Poincar\'e
  symmetries}}, \href{https://doi.org/10.1007/JHEP05(2020)071}{\emph{JHEP}
  {\bfseries 05} (2020) 071}
  [\href{https://arxiv.org/abs/2002.03865}{{\ttfamily 2002.03865}}].

\bibitem{Yin:2020oze}
C.~Yin and A.~Lucas, \emph{{Quantum operator growth bounds for kicked tops and
  semiclassical spin chains}},
  \href{https://doi.org/10.1103/PhysRevA.103.042414}{\emph{Phys. Rev. A}
  {\bfseries 103} (2021) 042414}
  [\href{https://arxiv.org/abs/2010.06592}{{\ttfamily 2010.06592}}].

\bibitem{Patramanis:2021lkx}
D.~Patramanis, \emph{{Probing the entanglement of operator growth}},
  \href{https://doi.org/10.1093/ptep/ptac081}{\emph{PTEP} {\bfseries 2022}
  (2022) 063A01} [\href{https://arxiv.org/abs/2111.03424}{{\ttfamily
  2111.03424}}].

\bibitem{Hornedal:2022pkc}
N.~H\"ornedal, N.~Carabba, A.S.~Matsoukas-Roubeas and A.~del Campo,
  \emph{{Ultimate Physical Limits to the Growth of Operator Complexity}},
  \href{https://arxiv.org/abs/2202.05006}{{\ttfamily 2202.05006}}.

\bibitem{Larkin1969QuasiclassicalMI}
A.I.~Larkin and Y.N.~Ovchinnikov, \emph{Quasiclassical method in the theory of
  superconductivity}, {\emph{Journal of Experimental and Theoretical Physics}
  (1969) }.

\bibitem{Roberts:2016hpo}
D.A.~Roberts and B.~Yoshida, \emph{{Chaos and complexity by design}},
  \href{https://doi.org/10.1007/JHEP04(2017)121}{\emph{JHEP} {\bfseries 04}
  (2017) 121} [\href{https://arxiv.org/abs/1610.04903}{{\ttfamily
  1610.04903}}].

\bibitem{Rabinovici:2020ryf}
E.~Rabinovici, A.~S\'anchez-Garrido, R.~Shir and J.~Sonner, \emph{{Operator
  complexity: a journey to the edge of Krylov space}},
  \href{https://doi.org/10.1007/JHEP06(2021)062}{\emph{JHEP} {\bfseries 06}
  (2021) 062} [\href{https://arxiv.org/abs/2009.01862}{{\ttfamily
  2009.01862}}].

\bibitem{Barbon:2019wsy}
J.L.F.~Barb\'on, E.~Rabinovici, R.~Shir and R.~Sinha, \emph{{On The Evolution
  Of Operator Complexity Beyond Scrambling}},
  \href{https://doi.org/10.1007/JHEP10(2019)264}{\emph{JHEP} {\bfseries 10}
  (2019) 264} [\href{https://arxiv.org/abs/1907.05393}{{\ttfamily
  1907.05393}}].

\bibitem{Dymarsky:2019elm}
A.~Dymarsky and A.~Gorsky, \emph{{Quantum chaos as delocalization in Krylov
  space}}, \href{https://doi.org/10.1103/PhysRevB.102.085137}{\emph{Phys. Rev.
  B} {\bfseries 102} (2020) 085137}
  [\href{https://arxiv.org/abs/1912.12227}{{\ttfamily 1912.12227}}].

\bibitem{Caputa:2021sib}
P.~Caputa, J.M.~Magan and D.~Patramanis, \emph{{Geometry of Krylov
  complexity}},
  \href{https://doi.org/10.1103/PhysRevResearch.4.013041}{\emph{Phys. Rev.
  Res.} {\bfseries 4} (2022) 013041}
  [\href{https://arxiv.org/abs/2109.03824}{{\ttfamily 2109.03824}}].

\bibitem{Balasubramanian:2022tpr}
V.~Balasubramanian, P.~Caputa, J.~Magan and Q.~Wu, \emph{{Quantum chaos and the
  complexity of spread of states}},
  \href{https://arxiv.org/abs/2202.06957}{{\ttfamily 2202.06957}}.

\bibitem{Chapman:2021jbh}
S.~Chapman and G.~Policastro, \emph{{Quantum computational complexity from
  quantum information to black holes and back}},
  \href{https://doi.org/10.1140/epjc/s10052-022-10037-1}{\emph{Eur. Phys. J. C}
  {\bfseries 82} (2022) 128}
  [\href{https://arxiv.org/abs/2110.14672}{{\ttfamily 2110.14672}}].

\bibitem{Lanczos1950AnIM}
C.~Lanczos, \emph{An iteration method for the solution of the eigenvalue
  problem of linear differential and integral operators}, {\emph{Journal of
  research of the National Bureau of Standards} {\bfseries 45} (1950) 255}.

\bibitem{Roberts:2018mnp}
D.A.~Roberts, D.~Stanford and A.~Streicher, \emph{{Operator growth in the SYK
  model}}, \href{https://doi.org/10.1007/JHEP06(2018)122}{\emph{JHEP}
  {\bfseries 06} (2018) 122}
  [\href{https://arxiv.org/abs/1802.02633}{{\ttfamily 1802.02633}}].

\bibitem{Jian:2020qpp}
S.-K.~Jian, B.~Swingle and Z.-Y.~Xian, \emph{{Complexity growth of operators in
  the SYK model and in JT gravity}},
  \href{https://doi.org/10.1007/JHEP03(2021)014}{\emph{JHEP} {\bfseries 03}
  (2021) 014} [\href{https://arxiv.org/abs/2008.12274}{{\ttfamily
  2008.12274}}].

\bibitem{Bhattacharjee:2022ave}
B.~Bhattacharjee, P.~Nandy and T.~Pathak, \emph{{Krylov complexity in large q
  and double-scaled SYK model}},
  \href{https://doi.org/10.1007/JHEP08(2023)099}{\emph{JHEP} {\bfseries 08}
  (2023) 099} [\href{https://arxiv.org/abs/2210.02474}{{\ttfamily
  2210.02474}}].

\bibitem{Bhattacharjee:2023uwx}
B.~Bhattacharjee, P.~Nandy and T.~Pathak, \emph{{Operator dynamics in
  Lindbladian SYK: a Krylov complexity perspective}},
  \href{https://arxiv.org/abs/2311.00753}{{\ttfamily 2311.00753}}.

\bibitem{PhysRevE.104.034112}
J.D.~Noh, \emph{Operator growth in the transverse-field ising spin chain with
  integrability-breaking longitudinal field},
  \href{https://doi.org/10.1103/PhysRevE.104.034112}{\emph{Phys. Rev. E}
  {\bfseries 104} (2021) 034112}.

\bibitem{Rabinovici:2021qqt}
E.~Rabinovici, A.~S\'anchez-Garrido, R.~Shir and J.~Sonner, \emph{{Krylov
  localization and suppression of complexity}},
  \href{https://doi.org/10.1007/JHEP03(2022)211}{\emph{JHEP} {\bfseries 03}
  (2022) 211} [\href{https://arxiv.org/abs/2112.12128}{{\ttfamily
  2112.12128}}].

\bibitem{Rabinovici:2022beu}
E.~Rabinovici, A.~S\'anchez-Garrido, R.~Shir and J.~Sonner, \emph{{Krylov
  complexity from integrability to chaos}},
  \href{https://doi.org/10.1007/JHEP07(2022)151}{\emph{JHEP} {\bfseries 07}
  (2022) 151} [\href{https://arxiv.org/abs/2207.07701}{{\ttfamily
  2207.07701}}].

\bibitem{Caputa:2021ori}
P.~Caputa and S.~Datta, \emph{{Operator growth in 2d CFT}},
  \href{https://doi.org/10.1007/JHEP12(2021)188}{\emph{JHEP} {\bfseries 12}
  (2021) 188} [\href{https://arxiv.org/abs/2110.10519}{{\ttfamily
  2110.10519}}].

\bibitem{Dymarsky:2021bjq}
A.~Dymarsky and M.~Smolkin, \emph{{Krylov complexity in conformal field
  theory}}, \href{https://doi.org/10.1103/PhysRevD.104.L081702}{\emph{Phys.
  Rev. D} {\bfseries 104} (2021) L081702}
  [\href{https://arxiv.org/abs/2104.09514}{{\ttfamily 2104.09514}}].

\bibitem{Kundu:2023hbk}
A.~Kundu, V.~Malvimat and R.~Sinha, \emph{{State dependence of Krylov
  complexity in 2d CFTs}},
  \href{https://doi.org/10.1007/JHEP09(2023)011}{\emph{JHEP} {\bfseries 09}
  (2023) 011} [\href{https://arxiv.org/abs/2303.03426}{{\ttfamily
  2303.03426}}].

\bibitem{Bender:2013qta}
C.M.~Bender and M.~Gianfreda, \emph{{Twofold Transition in PT-Symmetric Coupled
  Oscillators}}, \href{https://doi.org/10.1103/PhysRevA.88.062111}{\emph{Phys.
  Rev. A} {\bfseries 88} (2013) 062111}
  [\href{https://arxiv.org/abs/1305.7107}{{\ttfamily 1305.7107}}].

\bibitem{Peng:2014idi}
B.~Peng, c.K.~\"Ozdemir, F.~Lei, F.~Monifi, M.~Gianfreda, G.L.~Long et~al.,
  \emph{{Parity\textendash{}time-symmetric whispering-gallery microcavities}},
  \href{https://doi.org/10.1038/nphys2927}{\emph{Nature Phys.} {\bfseries 10}
  (2014) 394}.

\bibitem{PhysRev.38.815}
H.~Bateman, \emph{On dissipative systems and related variational principles},
  \href{https://doi.org/10.1103/PhysRev.38.815}{\emph{Phys. Rev.} {\bfseries
  38} (1931) 815}.

\bibitem{El-Ganainy:2018ksn}
R.~El-Ganainy, K.G.~Makris, M.~Khajavikhan, Z.H.~Musslimani, S.~Rotter and
  D.N.~Christodoulides, \emph{{Non-Hermitian physics and PT symmetry}},
  \href{https://doi.org/10.1038/nphys4323}{\emph{Nature Phys.} {\bfseries 14}
  (2018) 11}.

\bibitem{Ozdemir2019}
S.~Ozdemir, S.~Rotter, F.~Nori and L.~Yang, \emph{Parity–time symmetry and
  exceptional points in photonics},
  \href{https://doi.org/10.1038/s41563-019-0304-9}{\emph{Nature Materials}
  {\bfseries 18} (2019) 1}.

\bibitem{Ashida:2020dkc}
Y.~Ashida, Z.~Gong and M.~Ueda, \emph{{Non-Hermitian physics}},
  \href{https://doi.org/10.1080/00018732.2021.1876991}{\emph{Adv. Phys.}
  {\bfseries 69} (2021) 249}
  [\href{https://arxiv.org/abs/2006.01837}{{\ttfamily 2006.01837}}].

\bibitem{Haque:2022ncl}
S.S.~Haque, J.~Murugan, M.~Tladi and H.J.R.~Van~Zyl, \emph{{Krylov Complexity
  for Jacobi Coherent States}},
  \href{https://arxiv.org/abs/2212.13758}{{\ttfamily 2212.13758}}.

\bibitem{Perelomov}
A.M.~Perelomov, \emph{{Coherent states for arbitrary Lie group}},
  \href{https://doi.org/cmp/1103858078}{\emph{Communications in Mathematical
  Physics} {\bfseries 26} (1972) 222 }.

\bibitem{Gazeau:2009zz}
J.-P.~Gazeau, \emph{{Coherent states in quantum physics}} (2009).

\bibitem{Provost:1980nc}
J.P.~Provost and G.~Vallee, \emph{{RIEMANNIAN STRUCTURE ON MANIFOLDS OF QUANTUM
  STATES}}, \href{https://doi.org/10.1007/BF02193559}{\emph{Commun. Math.
  Phys.} {\bfseries 76} (1980) 289}.

\bibitem{Ashtekar:1997ud}
A.~Ashtekar and T.A.~Schilling, \emph{{Geometrical formulation of quantum
  mechanics}},  \href{https://arxiv.org/abs/gr-qc/9706069}{{\ttfamily
  gr-qc/9706069}}.

\bibitem{Brody:1999cw}
D.C.~Brody and L.P.~Hughston, \emph{{Geometric quantum mechanics}},
  \href{https://doi.org/10.1016/S0393-0440(00)00052-8}{\emph{J. Geom. Phys.}
  {\bfseries 38} (2001) 19}
  [\href{https://arxiv.org/abs/quant-ph/9906086}{{\ttfamily
  quant-ph/9906086}}].

\bibitem{Chattopadhyay:2023fob}
A.~Chattopadhyay, A.~Mitra and H.J.R.~van Zyl, \emph{{Spread complexity as
  classical dilaton solutions}},
  \href{https://doi.org/10.1103/PhysRevD.108.025013}{\emph{Phys. Rev. D}
  {\bfseries 108} (2023) 025013}
  [\href{https://arxiv.org/abs/2302.10489}{{\ttfamily 2302.10489}}].

\bibitem{osti_6577304}
H.~Feshbach and Y.~Tikochinsky, \emph{Quantization of the damped harmonic
  oscillator},
  \href{https://doi.org/10.1111/j.2164-0947.1977.tb02946.x}{\emph{Trans. N.Y.
  Acad. Sci.; (United States)} {\bfseries 38} (1977) }.

\bibitem{PhysRevA.88.062111}
C.M.~Bender, M.~Gianfreda, i.m.c.K.~\"Ozdemir, B.~Peng and L.~Yang,
  \emph{Twofold transition in $\mathcal{PT}$-symmetric coupled oscillators},
  \href{https://doi.org/10.1103/PhysRevA.88.062111}{\emph{Phys. Rev. A}
  {\bfseries 88} (2013) 062111}.

\bibitem{Deguchi:2018otx}
S.~Deguchi, Y.~Fujiwara and K.~Nakano, \emph{{Two quantization approaches to
  the Bateman oscillator model}},
  \href{https://doi.org/10.1016/j.aop.2019.02.004}{\emph{Annals Phys.}
  {\bfseries 403} (2019) 34}
  [\href{https://arxiv.org/abs/1807.04403}{{\ttfamily 1807.04403}}].

\bibitem{Deguchi:2019laq}
S.~Deguchi and Y.~Fujiwara, \emph{{Square-integrable eigenfunctions in
  quantizing the Bateman oscillator model}},
  \href{https://arxiv.org/abs/1910.08271}{{\ttfamily 1910.08271}}.

\bibitem{Bagarello:2019yag}
F.~Bagarello, F.~Gargano and F.~Roccati, \emph{{A no-go result for the quantum
  damped harmonic oscillator}},
  \href{https://doi.org/10.1016/j.physleta.2019.06.022}{\emph{Phys. Lett. A}
  {\bfseries 383} (2019) 2836}
  [\href{https://arxiv.org/abs/1906.05121}{{\ttfamily 1906.05121}}].

\bibitem{Bagarello2020}
F.~Bagarello, F.~Gargano and F.~Roccati, \emph{Some remarks on few recent
  results on the damped quantum harmonic oscillator},
  \href{https://doi.org/10.1016/j.aop.2020.168091}{\emph{Annals of Physics}
  {\bfseries 414} (2020) 168091}.

\bibitem{Fernandez2020}
F.M.~Fern'andez, \emph{Algebraic treatment of the bateman hamiltonian},
  {\emph{Canadian Journal of Physics} (2020) }.

\bibitem{Caputa:2022eye}
P.~Caputa and S.~Liu, \emph{{Quantum complexity and topological phases of
  matter}},  \href{https://arxiv.org/abs/2205.05688}{{\ttfamily 2205.05688}}.

\bibitem{Caputa:2022yju}
P.~Caputa, N.~Gupta, S.S.~Haque, S.~Liu, J.~Murugan and H.J.R.~Van~Zyl,
  \emph{{Spread Complexity and Topological Transitions in the Kitaev Chain}},
  \href{https://arxiv.org/abs/2208.06311}{{\ttfamily 2208.06311}}.

\bibitem{Erdmenger:2023wjg}
J.~Erdmenger, S.-K.~Jian and Z.-Y.~Xian, \emph{{Universal chaotic dynamics from
  Krylov space}}, \href{https://doi.org/10.1007/JHEP08(2023)176}{\emph{JHEP}
  {\bfseries 08} (2023) 176}
  [\href{https://arxiv.org/abs/2303.12151}{{\ttfamily 2303.12151}}].

\bibitem{Dehesa:1981}
J.~Dehesa, \emph{{Lanczos Method of Tridiagonalization, Jacobi Matrices and
  Physics}},
  \href{https://doi.org/https://doi.org/10.1016/S0034-4877(23)00059-9}{\emph{J.
  Comp. App. Math.} {\bfseries 7} (1981) 249}.

\bibitem{Muck:2022xfc}
W.~M\"uck and Y.~Yang, \emph{{Krylov complexity and orthogonal polynomials}},
  \href{https://doi.org/10.1016/j.nuclphysb.2022.115948}{\emph{Nucl. Phys. B}
  {\bfseries 984} (2022) 115948}
  [\href{https://arxiv.org/abs/2205.12815}{{\ttfamily 2205.12815}}].

\bibitem{deAlfaro:1976vlx}
V.~de~Alfaro, S.~Fubini and G.~Furlan, \emph{{Conformal Invariance in Quantum
  Mechanics}}, \href{https://doi.org/10.1007/BF02785666}{\emph{Nuovo Cim. A}
  {\bfseries 34} (1976) 569}.

\bibitem{viswanath2008recursion}
V.~Viswanath and G.~M{\"u}ller, \emph{The recursion method: application to
  many-body dynamics}, vol.~23, Springer Science \& Business Media (2008).

\bibitem{TaylorSeries.jl-2019}
L.~Benet and D.P.~Sanders, \emph{{TaylorSeries}.jl: Taylor expansions in one
  and several variables in julia}, {\emph{Journal of Open Source Software} }.

\bibitem{rackauckas2017differentialequations}
C.~Rackauckas and Q.~Nie, \emph{Differential{E}quations.jl--a performant and
  feature-rich ecosystem for solving differential equations in {J}ulia},
  {\emph{Journal of Open Research Software} {\bfseries 5} (2017) }.

\end{thebibliography}\endgroup

\appendix

\section{BCH formulae}

As is usual in computations of spread complexity using Hamiltonians selected from some symmetry algebra, coherent states are a useful tool to utilise.  We can use this, for example, to compute the return amplitude in an efficient way.  \\ \\
To this end, note the following finite dimensional faithful $4\times 4$ matrix representation of the operators quadratic in creation and annihilation operators
\begin{eqnarray}
\frac{\alpha}{2}a^\dag a^\dag + \frac{\alpha^*}{2}a a + \frac{\gamma}{4} \left( a^\dag a + a a^\dag \right) & = & \left( 
  \begin{array}{cccc} \frac{\gamma}{2} & i \alpha & 0 & 0 \\ 
  i \alpha^* & -\frac{\gamma}{2} & 0 & 0 \\ 0 & 0 & 0 & 0 \\ 0 & 0 & 0 & 0\end{array} \right)  \nonumber \\
  \frac{\alpha}{2}b^\dag b^\dag + \frac{\alpha^*}{2}b b + \frac{\gamma}{4} \left( b^\dag b + b b^\dag \right) & = & \left( 
  \begin{array}{cccc} 0 & 0 & 0 & 0 \\ 0 & 0 & 0 & 0 \\ 0 & 0 & \frac{\gamma}{2} & i \alpha \\ 
  0 & 0 & i \alpha^* & -\frac{\gamma}{2}  \end{array} \right)    \nonumber \\
  \rho a^\dag b^\dag + \rho^* a b & = & \left(  \begin{array}{cccc}  0 & 0 & 0 & i \rho  \\
  0 & 0 & i \rho^* & 0 \\
  0 & i \rho & 0 & 0 \\
  i \rho^* & 0 & 0 & 0 
 \end{array}  \right)   \nonumber  \\
 \rho a^\dag b + \rho^* a b^\dag & = & \left(  \begin{array}{cccc}  0 & 0 & \rho & 0  \\
  0 & 0 & 0 & -\rho^* \\
  \rho^* & 0 & 0 & 0 \\
  0 & -\rho & 0 & 0 
 \end{array}  \right)  \,.   \label{fourByfourRep}
\end{eqnarray}
Using this representation one can readily compute BCH formulas that are useful for computing the return amplitude.  Take, for example, the reference state
\begin{equation}
|\phi_r\rangle = |0,0\rangle \,.
\end{equation}
The BCH formulas allow one to decompose a group element into separate exponential of the generators.  For this choice of generators the following decomposition is useful
\begin{equation}
U = e^{\frac{z_{aa}}{2} a^\dag a^\dag  } e^{\frac{z_{bb}}{2} b^\dag b^\dag  } e^{\frac{z_{ab}}{2} a^\dag b^\dag  } e^{ \zeta 
 b^\dag a} e^{\Delta_a (a^\dag a + a a^\dag)} e^{\Delta_b (a^\dag a + a a^\dag)}   e^{\zeta' a^\dag b} e^{\frac{z'_{ab}}{2} a b  } e^{\frac{z'_{bb}}{2} b b  } e^{\frac{z'_{aa}}{2} a a  }  \,.\label{BCHgen}
\end{equation}
Note that the parameters above are complex, but not all independent.  The group is parametrised by 10 real parameters while the right-hand side of (\ref{BCHgen}) contains 20 real parameters.  Half of these are fixed by requiring that $U U^\dag = I$.   When acting on the reference state, the action of the group element (when decomposed in this way) simplifies greatly
\begin{eqnarray}
U |\phi_r\rangle & = & e^{\Delta_a + \Delta_b}   
 e^{\frac{z_{aa}}{2} A^\dag A^\dag} e^{\frac{z_{bb}}{2} B^\dag B^\dag} e^{ z_{ab} A^\dag B^\dag}  |0,0\rangle \nonumber \\
 & = & \mathcal{N} e^{\frac{z_{aa}}{2} A^\dag A^\dag} e^{\frac{z_{bb}}{2} B^\dag B^\dag} e^{ z_{ab} A^\dag B^\dag}  |0,0\rangle \,,
\end{eqnarray}
where $\mathcal{N}$ is a normalisation factor. These are precisely the generalized coherent states for the group \cite{Perelomov} and each coherent state is in a one-to-one correspondence with elements of the factor group $G / H$ where $H$ is the stationary subgroup associated with our choice of reference state.  Explicitly, these are the group elements 
\begin{equation}
e^{i c_a (a^\dag a + a a^\dag ) + i c_b (b^\dag b + b b^\dag ) + \zeta a b^\dag - \bar{\zeta} b a^\dag} \in H\,,
\end{equation}
which act trivially on the reference state.  \\ \\
The above discussion holds for an arbitrary element of the group, though we are primarily interested in decomposing the time-evolution operator in this way. Specifically, we want to compute the return amplitude
\begin{equation}
\langle \phi_r|  e^{- i t H} |\phi_r\rangle = e^{\Delta_a(t) + \Delta_b(t)}\,.
\end{equation}
The task of solving for the return amplitude is thus reduced to solving for the coefficients $\Delta_a, \Delta_b$ in the BCH decomposition of the time-evolution operator. In the infinite-dimensional representation (in terms of the creation and annihilation operators) this can be a tricky task but it is straightforward in the four-by-four matrix representation (\ref{fourByfourRep}). Since this is a faithful representation, it is sufficient in order to compute group multiplication identities such as (\ref{BCHgen}). Written in the four-by-four matrix representation, the right-hand side of (\ref{BCHgen}) becomes
\begin{equation}
U = \left( \begin{array}{cccc} 1 & i z_{aa} & 0 & i(z_{ab} - z_{aa} \zeta)  \\ 0 & 1 & 0 & -\zeta \\ \zeta & i z_{ab} & 1 & i(z_{bb} - z_{ab}\zeta) \\ 0 & 0 & 0 & 1 \end{array} \right) \left( \begin{array}{cccc}  e^{2 \Delta_a} & 0 & 0 & 0 \\ 0 &  e^{-2 \Delta_a} & 0 & 0 \\ 0 & 0 & e^{2 \Delta_b} & 0 \\ 0 & 0 & 0 & e^{-2 \Delta_b} \end{array}  \right) \left( \begin{array}{cccc}  1 & 0 & \zeta' & 0    \nonumber \\
i z_{aa}' & 1 & i z_{ab}' & 0  
  \nonumber \\
  0 & 0 & 1 & 0 \\ i(z_{ab}' - z_{aa} \zeta') & -\zeta' & i(z_{bb}' - z_{ab}' \zeta' ) & 1 \end{array}    \right)
\end{equation}
Using usual matrix exponentiation one can compute the time-evolution operator for any choice of Hamiltonian from the symmetry algebra, though we have suppressed the explicit expressions for brevity.  It is a straightforward exercise to solve for the coefficients on the right-hand side of the expression and thus compute the return amplitude.  Indeed, using this same procedure, one can readily compute the return amplitude with any coherent state as reference state.  

\section{Spread complexity for a single oscillator}

Some of the features of spread complexity for the Bateman oscillator have parallels in a much simpler system - that of a single oscillator.  The Hamiltonian in this case is an element of $su(1,1)$.  Working in the representation of creation and annihilation operators, a general Hamiltonian can be expressed as 
\begin{equation}
H = \frac{\gamma}{4}( a^\dag a + a a^\dag) + \frac{\alpha}{2} a^\dag a^\dag + \frac{\alpha^*}{2} a a\,,
\end{equation}
and the manifold of generalized coherent states is coordinatized by
\begin{equation}
|z) = e^{\frac{z}{2} a^\dag a^\dag} |0\rangle \ \ \ ; \ \ \ |z|^2 < 1
\end{equation}
The operators from this algebra can be classified into three different classes \cite{deAlfaro:1976vlx}
depending on the sign of
\begin{equation}
\Delta = \gamma^2 - 4\alpha \alpha^*
\end{equation}
If $\Delta$ is positive the eigenfunctions are normalizable (i.e. square integrable).  If $\Delta < 0$ the spectrum is not bounded from below, if $\Delta=0$ the spectrum is continuous and if $\Delta > 0$ the spectrum is discrete.  These are classified as hyperbolic, parabolic and elliptic generators respectively \cite{deAlfaro:1976vlx}.  An important feature is that Hamiltonians related to one another by a unitary transformation have the \textit{same} value for $\Delta$. This can be checked explicitly by performing a general Bogoliubov transformation on the operators
\begin{equation}
a \rightarrow a \cosh(\rho) e^{i \phi} + a^\dag \sinh(\rho) e^{i \phi}
\end{equation}
and computing the value for $\Delta$ after the transformation.  Indeed, the value for $\Delta$ is precisely the value that dictates the time-dependence of spread complexity.  The spread complexity of the time-evolved vacuum state (starting form the vacuum as reference state) is given by  
\begin{equation}
C(e^{-i t H}|0\rangle; H, |0\rangle) = \frac{2 |\alpha|^2}{4 |\alpha|^2 - \gamma^2} \sinh^2(\sqrt{4 |\alpha|^2 - \gamma^2} t)
\end{equation}
Note that the elliptic generators will give rise to a periodic spread complexity.  Since the value of $\Delta$ is not affected by unitary transformations we can immediately conclude that an elliptic generator acting on \textit{any} $su(1,1)$ coherent state will give rise to periodic spread complexity.  As representative of the elliptic generators we can pick $H_e = \omega a^\dag a$ and consider
\begin{eqnarray}
R(t) & = & (1 - \bar{z} z)^{-1} \langle e^{\frac{\bar{z}}{2} a a} e^{- i t H_e} e^{\frac{z}{2} a^\dag a^\dag} |0\rangle    \nonumber \\
& = & \frac{1 - \bar{z} z e^{-i \omega t}}{1 - \bar{z} z}
\end{eqnarray}
for which the spread complexity is 
\begin{equation}
C(t) = \frac{4 |z|^2 }{(1 - |z|^2)^2} \sin^2(\omega t)
\end{equation}
Evidently, the choice of reference state can affect the overall factor multiplying the spread complexity, but cannot change the fact that the elliptic generator gives rise to spread complexity that is periodic.

\section{Numerical analysis}

In this appendix we discuss the numerical implementation of the linearly interacting coupled oscillator system, and detail the intricacies of computing the Lanczos coefficients and Spread Complexity. Analytic parts of the results were obtained in \texttt{Mathematica} while \texttt{Julia} was chosen for the numerical results for its well-professed computational power.\\

\noindent
As detailed in the main text, Krylov/State Complexity is computed as a weighted sum of the complexity wavefunction probabilities. These, in turn, are determined through two distinct sets of Lanczos coefficients, ${  \left\{ a_{n} \right\} }$ and ${  \left\{ b_{n} \right\} }$, where ${ n \in [0,D] }$ with $D$ the dimension of the Krylov space generated by the Lanczos (or any other appropriate) algorithm. There are two standard approaches to computing the Lanczos coefficients; through the Krylov basis vectors and through the return amplitude. The former has already been summarized in Section 3. We provide a brief summary of the latter - also sometimes called the  {\it moments method} - here.

\subsection{Lanczos coefficient computation via return amplitude}
The Lanczos coefficients can be obtained via the moments of the return amplitude as follows: Given a return amplitude ${ G(t) }$, the ${ n^{\textnormal{th}} }$ moment of ${ G(t) }$ is defined as, 
\begin{equation}\label{eq:momentsofretamp}
	m_{n} = \frac{1}{G(0)}\left( -i ~ \frac{d}{dt} \right)^{n}\ G(t) \big|_{t=0}\,.
\end{equation}
The Lanczos ${ b_{n} }$ coefficients can then be computed using a Hankel transform of the moment matrix, ${ M_{ij} = m_{i+j} }$ as follows, 
\[
\begin{aligned}
		b_{1}^{2n} b_{2}^{2(n-1)} \dots{} b_{n}^{2} = \det (M_{ij}) \big|_{0 \leq i,j \leq n }\,.
\end{aligned}
\]
The content of this statement is that the ${ b_{n} }$'s are obtained by choosing successively larger values of $n$. Starting with $n=1$, we find ${ b_{1}^{2} = \det (M_{ij}) \big|_{0 \leq i,j \leq 1 } }$. This is then substituted into the $n=2$ relation ${ b_{1}^{4} \cdot b_{2}^{2} = \det (M_{ij}) \big|_{0 \leq i,j \leq 2 } }$ to compute $b_{2}$, and so on. For more details on the above, as well as the computation of ${ a_{n} }$'s, we refer the interested reader to \cite{viswanath2008recursion}.

\noindent
The moments method offers a neat way to compute the Lanczos coefficients analytically and is straightforward to implement numerically. However, it suffers from numerical instabilities and must be implemented with care taken to use the appropriate precision. \texttt{Julia} happens to be well-suited to the task. Most of the numerical computational time is spent computing the determinant of the moment matrix for each ${ n }$; hence it is recommended to pre-process (lower-upper decomposition etc.) the matrix before computing its determinant. Due to the large size of the moment matrix with increasing ${ n }$, the determinant computation determines the maximum number of coefficients one can obtain in a reasonable time frame. As an example, it takes the code approximately 5 minutes to compute the first 200 coefficients in \texttt{Julia} via the moments method. \\

\noindent
As a general rule, the GRA method (see sec.(\ref{subsec:GRA&SOE}) performs better when fewer coefficients are required as compared to the SOE method whereas, when a larger number of coefficients is required, numerical instabilities in the moments method propagate and grow too rapidly for the method to be reliable even while enforcing very high numerical precision. A simple way to understand this would be to consider \eqref{eq:momentsofretamp}. The moments method involves computing the ${ n^{\textnormal{th}} }$ order derivative of the return amplitude, which will be proportional to ${ n! }$, to obtain the ${ n^{\textnormal{th}} }$ coefficient. For ${ b_{100} }$ and ${ a_{100} }$, for example, one would need to compute the ${ 100^{\textnormal{th}} }$ derivative which is proportional to ${ 100! \sim 10^{157} }$. Then moments method involves taking determinants of the moment matrices for each ${ n }$, containing these large values, and a ratio of these determinants. This puts any numerical implementation of the moments method at a disadvantage against the Krylov vector method. In spite of this caveat, we found it possible to efficiently obtain up to ${ 200 }$ coefficients using the moment method and high-precision computation, although this statement is contingent on how complicated the form of the return amplitude is. Pragmatically, what is of most importance in these computations is the time up to which the complexity can be accurately approximated with a particular choice of the artificial cut-off, $N$, of the Krylov space dimension.\\

\begin{figure}[ h ]
	\centering
	\begin{subfigure}{0.3\linewidth}
		\includegraphics[trim={0 0 0 1cm},clip,width=\linewidth]{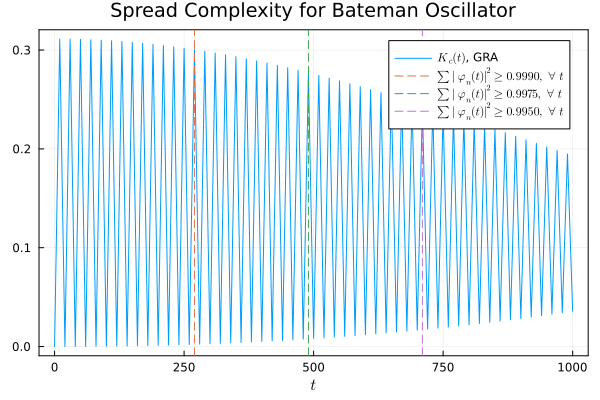}
		\caption{${ N=10 }$}
		\label{fig:ud&weakN10γ0.0001}
	\end{subfigure}
	\hfill
	\begin{subfigure}{0.3\linewidth}
		\includegraphics[trim={0 0 0 1cm},clip,width=\linewidth]{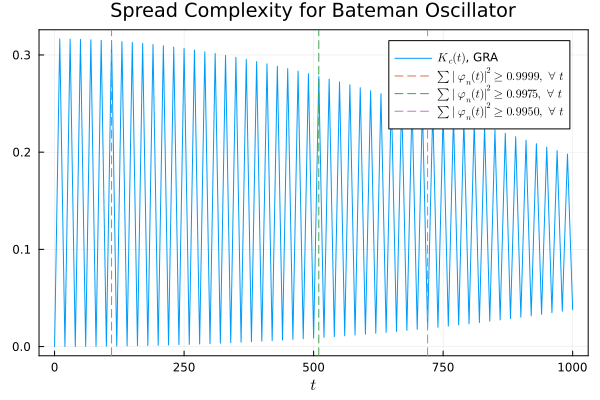}
		\caption{${ N=30 }$}
		\label{fig:ud&weakN30γ0.0001}
	\end{subfigure}
	\hfill
	\begin{subfigure}{0.3\linewidth}
		\includegraphics[trim={0 0 0 1cm},clip,width=\linewidth]{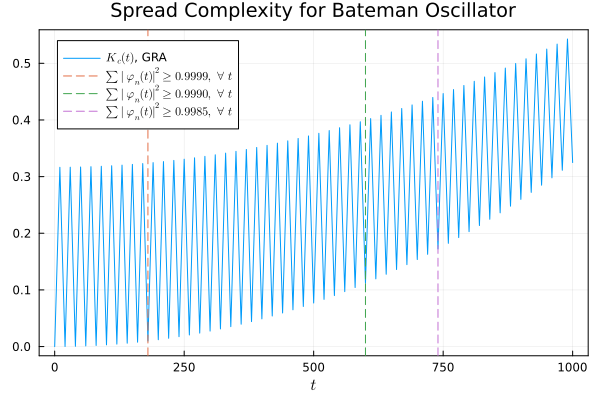}
		\caption{${ N=50 }$}
		\label{fig:ud&weakN50γ0.0001}
	\end{subfigure}
	\vspace{1cm}\vfill
	\begin{subfigure}{0.45\linewidth}
		\includegraphics[trim={0 0 0 1cm},clip,width=\linewidth]{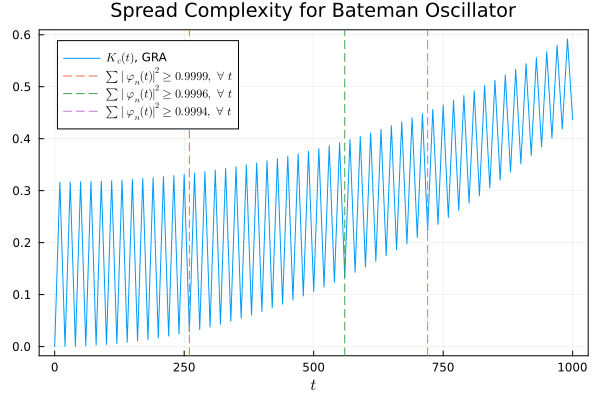}
		\caption{${ N=80 }$}
		\label{fig:ud&weakN80γ0.0001}
	\end{subfigure}
	\hfill
	\begin{subfigure}{0.45\linewidth}
		\includegraphics[trim={0 0 0 1cm},clip,width=\linewidth]{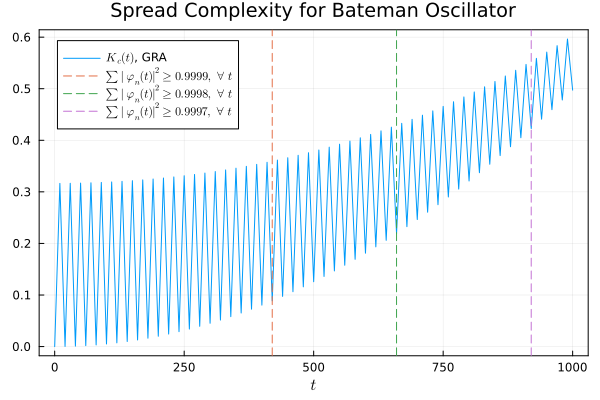}
		\caption{${ N=120 }$}
		\label{fig:ud&weakN120γ0.0001}
	\end{subfigure}
	
	\caption{A comparison of the complexity obtained through varying the artificial Krylov space cut-off ${ N }$ for the same choice of parameters : ${ \varkappa = 2.0,\ \epsilon = 0.0001414 < \epsilon_{1},\ \gamma = 0.0001,\ \Omega = 1.0}$ i.e. the system is under-damped and weakly coupled. These specific parameter choices have been made so that the strain between oscillations and linear growth is apparent. It is clear that with smaller values of ${ N }$ one may miss important features even though most of the total probability, ${ \geq 99.75\% }$, is captured. We note that with this set of parameters (and others presented in this appendix) it is sufficient to take approximately ${ N = 50 }$ to capture the physics, and is robust even for late times, which is small enough that using the Generated from Return Amplitude (GRA) method out-performs System of Equations (SOE) method significantly; see section C.2. Increasing ${ N }$ allows one to probe for progressively smaller time difference and later times more accurately (\textit{i.e.}, with lower tolerance for fault/a higher percentage of total probability captured).}
	\label{fig:NComp}
\end{figure}

\begin{figure}[ h ]
	\centering
	\begin{subfigure}{0.45\linewidth}
		\includegraphics[trim={0 0 0 1cm},clip,width=\linewidth]{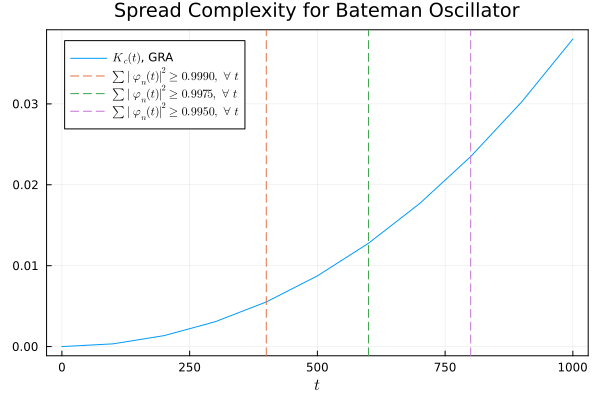}
		\caption{${ N=30 }$}
		\label{fig:N30no-osc}
	\end{subfigure}
	\hfill
	\begin{subfigure}{0.45\linewidth}
		\includegraphics[trim={0 0 0 1cm},clip,width=\linewidth]{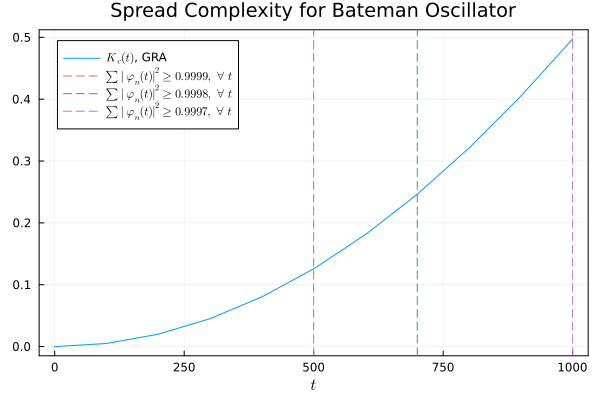}
		\caption{${ N=120 }$}
		\label{fig:N50no-osc}
	\end{subfigure}
	\caption{This figure serves to demonstrate that one should carefully choose the time intervals to probe. The parameters here are the same as in Fig.(\ref{fig:NComp}) but there is no oscillation to be observed even with ${ N=120 }$ because the smallest time interval probed is ${ \Delta t = 50.0 }$ compared to ${ \Delta t = 1.0 }$ for Fig.(\ref{fig:NComp}) (and the other figures in this appendix). We observe that with increasing ${ N }$, ${ \frac{\Delta t}{t_{\textnormal{max.}}} }$, where ${ t_{\textnormal{max.}} }$ is the total time up to which the system is evolved, can be made smaller to obtain complexity with more accuracy i.e. we can probe the system for smaller time intervals and for longer times as ${ N }$ is increased.}	
	\label{fig:NCompNoOsc}
\end{figure}

\noindent
The utility of computing Lanczos coefficients is that they contain all the dynamical information about the system \cite{Rabinovici:2020ryf}. One may therefore be justifiably concerned that the moments method may not generate enough coefficients to probe the time regimes of interest to sufficient accuracy to capture the physics. However, our numerical analysis here demonstrates that, at least for this particular class of problems, by appropriately choosing the parameters of the system, even ${ N=30 }$ suffices to capture the expected physics -- allowing for some tolerance -- as benchmarked by our analytic computations in various tractable limiting cases (see Fig.(\ref{epsilon0p1Plot})). Figs.\ref{fig:NComp}, and \ref{fig:NCompNoOsc} show a comparison of the complexity with increasing ${ N }$.\\

\noindent
As has been noted, the moment method requires taking high order derivatives of the return amplitude. To improve the efficiency of this process, we found it best to power-series expand the return amplitude to compute high order derivatives rather than plugging the return amplitude into a generic derivative function; the latter often restricts one to taking only the first few derivatives for a sufficiently involved return amplitude - as in our case. While this trick is merely useful in finite-dimensional systems, it is indispensable in an infinite dimensional Krylov space where the numerical implementation necessarily needs an introduction of a cut-off in the dimension of the Krylov space. We utilized, \texttt{TaylorSeries.jl} \cite{TaylorSeries.jl-2019}, in our implementation.\\

\noindent
Finally, it is important to control the precision while implementing the moment method due to the inherent numerical instability of the technique. For finite-dimensional spaces, this instability may result in non-truncation of the Krylov space as successive coefficients are computed by taking ratios. If there is loss of precision, a coefficient that was supposed to be zero would instead have a small value but would result in the subsequent coefficient being very large; similar issues occur for infinite dimensional Krylov spaces. However, these loss-of-precision errors are easy to manage.

\subsection{Computing the complexity wavefunctions}\label{subsec:GRA&SOE}
Once the Lanczos coefficients are obtained, there are two straightforward methods to compute the complexity wavefunctions: 
\begin{itemize}
	\item System of Equations (SOE) : the standard method involves simultaneously solving the following system of equations,  
	
		\begin{equation}\label{eq:complexitySOE}
			\left(\begin{matrix}
			  \dot{\phi}_0(t) \\
			  \dot{\phi}_1(t) \\
			  \vdots \\
			  \dot{\phi}_{n}(t)
			\end{matrix}\right)
				= -i \left(\begin{matrix}
				  a_1 & b_1 & 0 & 0 ~~\cdots~~ 0 \\
				  b_1 & a_2 & b_2 & 0 ~~\cdots~~ 0 \\
				  0 & b_2 & a_3 & b_3 ~~\cdots~~ 0 \\
				  \vdots & \vdots & \vdots &\ddots\\
				  0 & 0 & 0 & \cdots ~~ a_{n}
			\end{matrix}\right) \times 
			\left(\begin{matrix} 
			  \phi_0(t) \\
			  \phi_1(t) \\
			  \vdots \\
			  \phi_{n}(t)
			\end{matrix}\right)
		,\end{equation}
		
with the boundary conditions : ${ \phi_{0}(0) = 1 }$ and ${ \phi_{n}(0) = 0 \ \forall\ n>0 }$. Crucially, here the system is restricted to  a finite-dimensional approximation of the full system by the cut-off $N$. All $N$ of the complexity wavefunctions are known for all times after solving (\ref{eq:complexitySOE}). Obviously, due to the introduction of cut-off, the solutions will not accurately represent the true complexity wavefunctions after a specific time.

	\item Generated from Return Amplitude (GRA) : one can also generate the complexity wavefunctions from the return amplitude, where the first wavefunction is simply the return amplitude, ${ G(t) = \phi_{0}(t) }$. The rest of the wavefunctions can then be generated successively using, rather than simultaneously solving, \eqref{eq:complexitySOE} along with the boundary conditions, ${ \phi_{0}(0) = 1 }$ and ${ \phi_{n}(0) = 0 \ \forall\ n>0 }$. The tridiagonal structure of this equation allows for efficiency improvements and makes this the preferred choice to SOE, when the dimension of, or cut-off on, Krylov space is small enough. The cut-off in this case amounts to the statement that we will restrict ourselves only to the computation of the first $N$ wavefunctions -- the higher valued wavefunctions still exist in this paradigm, we just do not compute them; we have not restricted ourselves to a finite-dimensional approximation of the system as was done in SOE. For SOE, choosing a finite cut-off implies that ${ \phi_{n>N}(t) = 0\ \forall\ t }$ - which is strictly only true at ${ t = 0 }$ but may practically be so for longer times but not all times -  while there is no such assumption in GRA. Hence, as time-evolves, one might expect that some fraction of the total probability will not be captured as it is lost to the wavefunctions $\phi_n(t)$ with $n>N$. 
	
\end{itemize}

\begin{figure}[ h ]
	\centering
	\begin{subfigure}{0.45\linewidth}
		\includegraphics[trim={0 0 0 1cm},clip,width=\linewidth]{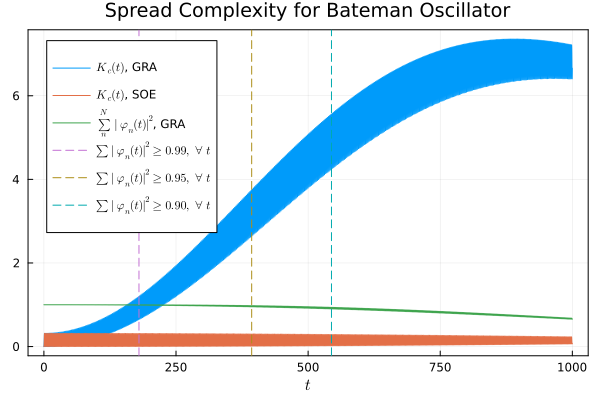}
		\caption{${ \gamma = 0.001,\ \epsilon = 0.001414 < \epsilon_{1}}$}
		\label{fig:ud&weakN50γ0.001}
	\end{subfigure}
	\hfill
	\begin{subfigure}{0.45\linewidth}
		\includegraphics[trim={0 0 0 1cm},clip,width=\linewidth]{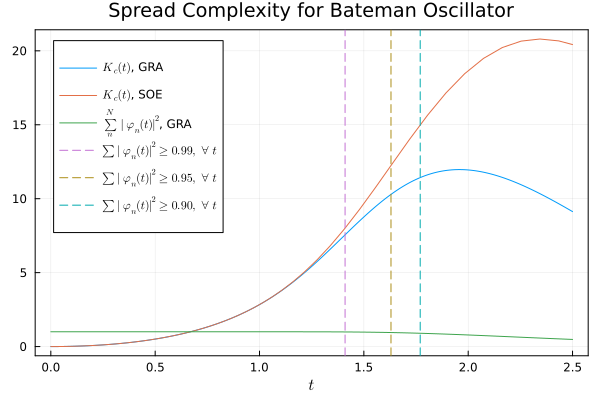}
		\caption{${ \gamma = 1.0,\ \epsilon = 1.414 < \epsilon_{1},\ \Delta t = 0.01}$}
		\label{fig:ud&weakγ1.0}
	\end{subfigure}
	
	\caption{A comparison of two methods for obtaining complexity : GRA and SOE. The parameters are chosen to be ${ \varkappa = 2.0,\ \Omega = 1.0,\ N = 50}$ such that the under-damped system is weakly coupled. We can see good agreement between the methods for early times in both (a) and (b), but the SOE method fails to capture the complexity growth at later times. However, in some instances it provides good approximations for intermediate times, as can be seen in (b). Nonetheless, the SOE method eventually fails in both instances. }
	\label{fig:GRAvsSOE}
\end{figure}

\noindent
The GRA method has two significant advantages over SOE method: (\textit{i}) it is orders of magnitude faster, and (\textit{ii}) the probability sum of the complexity wavefunctions, ${ \sum_{n=0}^{N} \left\lvert \phi_{n}(t) \right\rvert^{2} }$, provides a measure of the total probability captured by introducing the cut-off at ${ N }$ while the same probability sum for the SOE method demonstrates the accuracy of the differential equation solver with the cut-off ${ N }$. Consequently, it is not straightforward to characterize the error in the computed complexity via the SOE method, whereas the probability sum readily gives an error estimate for the GRA. The probability sum, computed via GRA,  is expected to be 1.0 for early times, even with small ${ N }$, but as the system evolves it is expected to decrease because ${ \phi_{N+1}(t>t_{0}) \neq 0}$, for some ${ t_{0} }$, whereas for the SOE method fixing the cut-off at ${ N }$ assumes that $\phi_{M>N}$ vanishes identically. The GRA method implicitly tracks the missing probabilities while the SOE method, by design, cannot. Naively, the error reliably computed in the GRA method serves as an approximation for the error of the SOE method, at least qualitatively. For the SOE, we used the \texttt{DifferentialEquations.jl} \cite{rackauckas2017differentialequations} differential equations solver package. The SOE and GRA results agree up to a certain time which, as one might expect, depends on the parameters that the return amplitude is computed from, as can be seen in fig.\ref{fig:GRAvsSOE}. Any generic simultaneous differential equation solver will need to identify the type of the system of equations for the SOE method to work well, as well as adaptively determine suitable parameter intervals while integrating the differential equations. This contributes significantly to the computational time for the SOE method. The GRA method is impervious to the structure that the equations form as a system and, consequently, is more optimized.\\ 

\begin{figure}[ h ]
	\centering

    \begin{subfigure}{0.45\linewidth}
        \includegraphics[trim={0 0 0 1cm},clip,width = \linewidth]{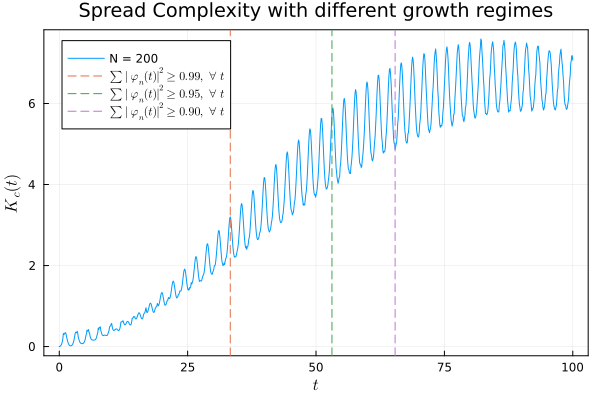}
        \caption{ ${ \varkappa = 2.0,\ \epsilon = 0.01414,\ \gamma = 0.01,\ \Omega = 1.0}$, ${N=200 }$.}
        \label{fig:ComplexityGrowthRegimesA}
    \end{subfigure}
    \hfill
    \begin{subfigure}{0.45\linewidth}
        \includegraphics[trim={0 0 0 1cm},clip,width = \linewidth]{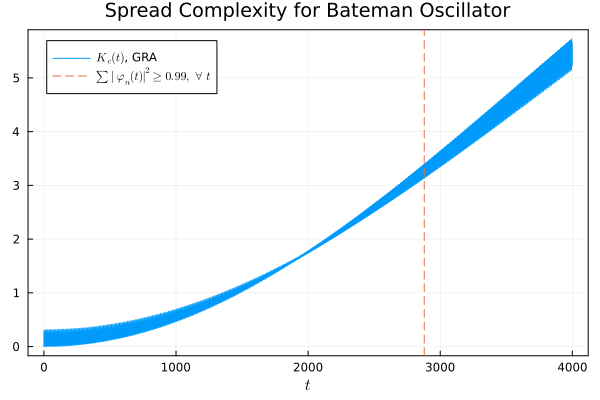}
        \caption{${ \varkappa = 2.0,\ \epsilon = 0.0001414,\ \gamma = 10^{-4},\ \Omega = 1.0}$, ${\ N=120 }$. }
        \label{fig:ComplexityGrowthRegimesB}
    \end{subfigure}
    
	\caption{The spread complexity showing two distinct growth regimes: linear-growth-dominant; and oscillation-dominant. The former is truly brought into contrast in the Fig.(\ref{fig:ComplexityGrowthRegimesB}) around t=2000 as the oscillations are heavily suppressed.}
	\label{fig:ComplexityGrowthRegimes}
\end{figure}

\begin{figure}[ h ]
	\centering
	
    \includegraphics[trim={0 0 0 1cm},clip,width = 0.6\textwidth]{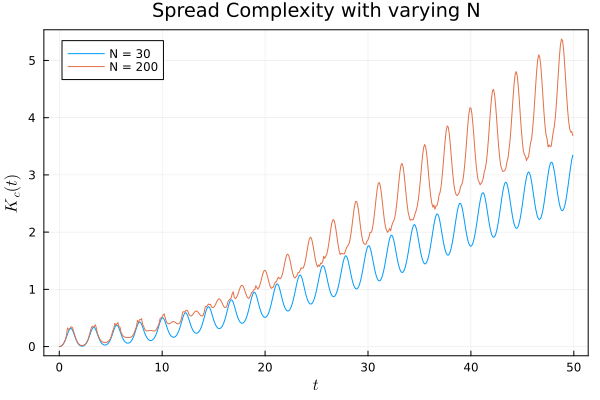}
	\caption{The spread complexity with parameters : ${ \varkappa = 2.0,\ \epsilon = 0.01414,\ \gamma = 0.01}$, and ${ \Omega = 1.0 }$ for different values of ${ N }$. There are linear-growth-dominant and oscillation-dominant regimes corresponding to Fig.(\ref{fig:ComplexityGrowthRegimesA}). With ${ N = 200 }$, there is a clear suppression of oscillations as compared to ${ N=30 }$. With any ${ N \leq 50 }$ the linear-growth-dominant regime in complexity will be indistinguishable since the Lanczos coefficients are mostly oscillation-dominant as mentioned in Fig.(\ref{fig:LancCoeffGrowthRegimes}).}
	\label{fig:ComplexityGrowthRegimesNComp}
\end{figure}

\begin{figure}[ h! ]
	\centering
	
    \includegraphics[width = 0.6\textwidth]{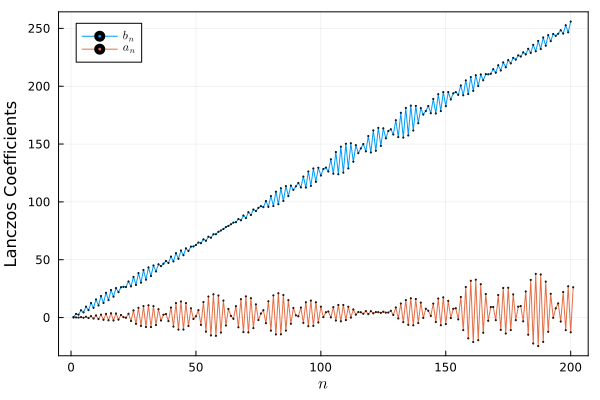}
	\caption{The first 200 Lanczos coefficients with parameters : ${ \varkappa = 2.0,\ \epsilon = 0.01414}$, ${ \gamma = 0.01, \text{ and } \Omega = 1.0 }$. There are both linear-growth-dominant and oscillation-dominant regimes corresponding to Figs.(\ref{fig:ComplexityGrowthRegimesA}), (\ref{fig:ComplexityGrowthRegimesNComp}). With ${ N \leq 50 }$ the linear-growth-dominant regime in complexity is absent since the Lanczos coefficients are mostly oscillation-dominant.}
	\label{fig:LancCoeffGrowthRegimes}
\end{figure}

\noindent
Finally, observe that for the case of the under-damped weakly coupled system, it is expected from our analytic analysis that spread complexity should oscillate with a linearly growing envelope. This is confirmed through numerical analysis in Fig.(\ref{epsilon0p1Plot}). The numerical analysis also reveals a more interesting feature upon further investigation. In Fig.(\ref{fig:ComplexityGrowthRegimes}), we observe that there is a brief regime of time around ${ t=17.0 }$ where the linear growth in complexity dominates the oscillations. For other choices of parameters we also observe that there is a periodic shift between oscillation-dominant and linear-growth-dominant regimes. This correlates with a corresponding behaviour in the associated Lanczos coefficients, as shown in Fig.(\ref{fig:LancCoeffGrowthRegimes}). This is natural since the Lanczos coefficients encode all the dynamical information of a system. Fig.(\ref{fig:ComplexityGrowthRegimesNComp}) shows a comparison of spread complexity for different values of ${ N} $. The linear-growth-dominant regime in complexity is not observed if ${ N }$ is chosen such that the corresponding regime is not captured in the Lanczos coefficients (more specifically, the ${ b_{n} }$s, as shown in Fig.(\ref{fig:LancCoeffGrowthRegimes}). We conjecture that the absence of linear-growth-dominant regimes is not due to taking a small ${ N }$, but is in fact a feature of the system rather than a numerical artefact. This feature is also unique to the specific case of under-damped weakly coupled oscillators. It would be interesting to test this in other, similar systems.

\end{document}